\documentclass[journal]{IEEEtran}

\ifCLASSINFOpdf
\else
\fi
\usepackage{graphicx}
\usepackage{amsmath}
\usepackage{amssymb}
\usepackage{multirow}
\usepackage{enumerate}
\usepackage{algorithm}
\usepackage{algorithmic}
\usepackage{slashbox}
\usepackage{array}
\usepackage{lineno,hyperref}
\usepackage{bm}
\usepackage{hyperref}
\usepackage{subfig}
\usepackage{epstopdf}
\newcommand{\tabincell}[2]{\begin{tabular}{@{}#1@{}}#2\end{tabular}}

\begin{document}

\title{Refining Image Categorization by Exploiting Web Images and General Corpus}

\author{Yazhou~Yao,~\IEEEmembership{Student Member,~IEEE,}
	Jian~Zhang,~\IEEEmembership{Senior Member,~IEEE,}
	Fumin~Shen,~\IEEEmembership{Member,~IEEE,}
	Xiansheng~Hua,~\IEEEmembership{Fellow,~IEEE,}
	Wankou Yang,~\IEEEmembership{Member,~IEEE,} 
	and~Zhenmin~Tang
	
	\thanks{Y. Yao and J. Zhang are with the Global Big Data Technologies Center, University of Technology Sydney, Australia.}
	\thanks{F. Shen is with the School of Computer Science and Engineering, University of Electronic Science and Technology of China.}
	\thanks{X. Hua is a researcher/senior director in Alibaba Group, China.}
	\thanks{W. Yang is with the School of Automation, SouthEast University, China}
	\thanks{Z. Tang is with the School of Computer Science and Engineering, Nanjing University of Science and Technology, China.}}

\markboth{IEEE Transactions on Multimedia}%
{Shell \MakeLowercase{\textit{et al.}}: Bare Demo of IEEEtran.cls for IEEE Journals}

\maketitle

\begin{abstract}
Studies show that refining real-world categories into semantic subcategories contributes to better image modeling and classification. Previous image sub-categorization work relying on labeled images and WordNet's hierarchy is not only labor-intensive, but also restricted to classify images into NOUN subcategories. To tackle these problems, in this work, we exploit general corpus information to automatically select and subsequently classify web images into semantic rich (sub-)categories. The following two major challenges are well studied: 1) noise in the labels of subcategories derived from the general corpus; 2) noise in the labels of images retrieved from the web. Specifically, we first obtain the semantic refinement subcategories from the text perspective and remove the noise by the relevance-based approach. To suppress the search error induced noisy images, we then formulate image selection and classifier learning as a multi-class multi-instance learning problem and propose to solve the employed problem by the cutting-plane algorithm. The experiments show significant performance gains by using the generated data of our way on both image categorization and sub-categorization tasks. The proposed approach also consistently outperforms existing weakly supervised and web-supervised approaches. 
\end{abstract}


\IEEEpeerreviewmaketitle

\section{Introduction}

\IEEEPARstart{I}mage categorization has achieved a great progress in the past few years, but it still needs a massive amount of manually labeled data \cite{yan2016image,chatzilari2016salic,bai2016multiple,wang2016csps,jian2016semi}. Meanwhile, image sub-categorization has been used to improve performance in a wide variety of vision tasks. For example, object detection \cite{felzenszwalb2010object}, animal behaviour analysis \cite{fox2009nonparametric} and image classification \cite{mansur2008improving}. Subdividing categories into subcategories multiples the number of labels, aggravating the annotation problem.

With the development of Internet, the number of digital images is growing extremely rapidly. How to effectively categorize these images has become an increasingly serious 
\begin{figure}[tbp]
	\centering
	\includegraphics[width=0.48\textwidth]{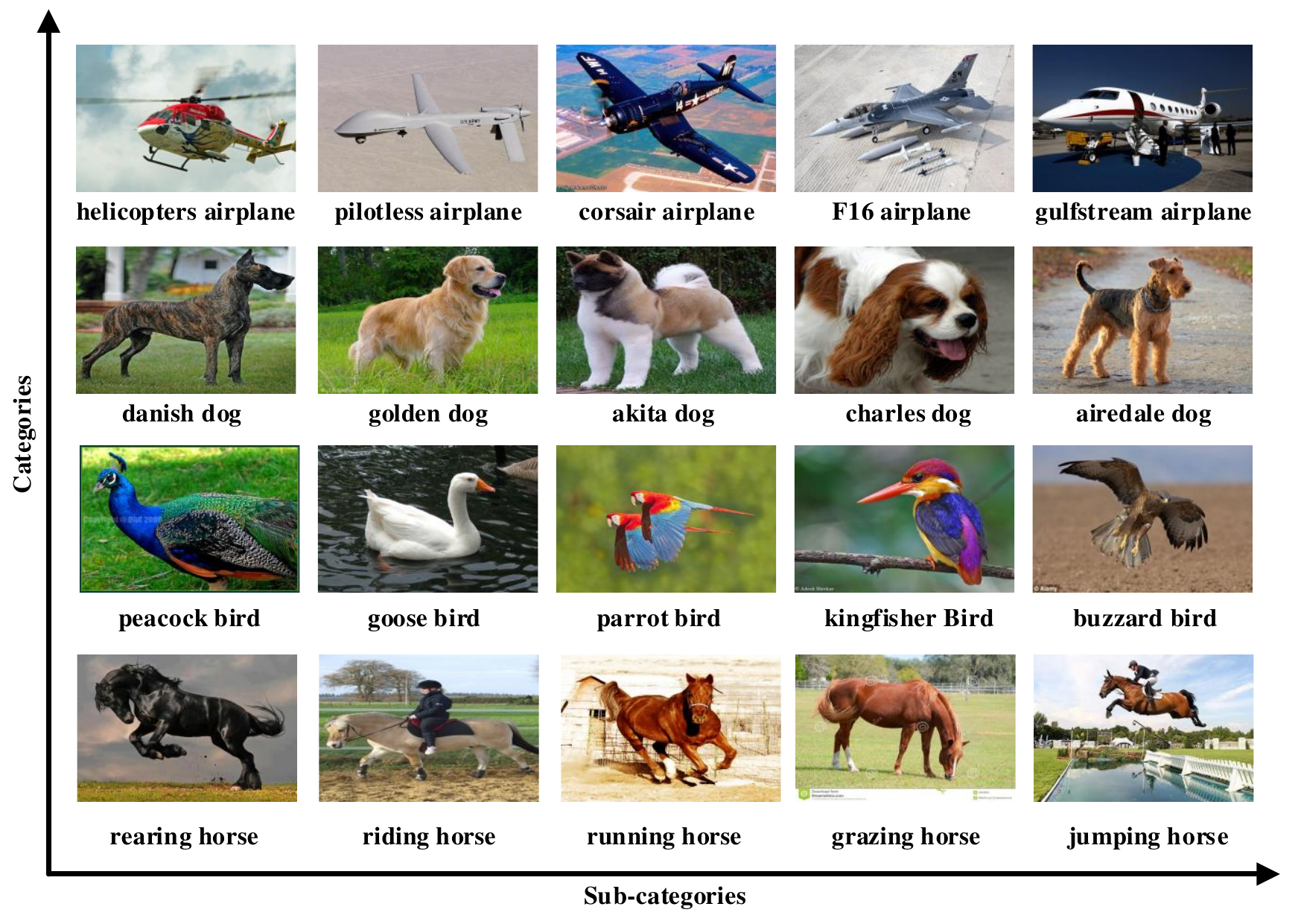}
	\caption{Image categorization and sub-categorization with the vertical axis representing different categories and the horizontal axis representing the different subcategories of the same category.}
	\label{fig1}  
\end{figure}
problem. Further, previously defined NOUN categories usually unable to have a good description for these emerging images, which have a variable appearance, positions, and poses \cite{hoai2013discriminative,ristin2015categories}. The categories need to be divided into more semantic rich subcategories to cover the finer semantic differences. ImageNet \cite{deng2009imagenet} is an image dataset organized according to the WordNet \cite{miller1995wordnet} hierarchy. It provides the research community not only with thousands of categories and millions of images, but also with refinement labels in a hierarchy. However, the process of constructing ImageNet is both time-consuming and labor-intensive. For example, it has taken several years to construct the ImageNet. In addition, the ImageNet requires the pre-existing expert knowledge WordNet and it only contains the NOUN subcategories (e.g., category ``airplane", ``dog" and ``bird"  in Fig. \ref{fig1}), it does not contain the VERB subcategories (e.g., category ``horse" in Fig. \ref{fig1}).

To reduce the cost of manual annotation, automatic methods by exploiting web images for image categorization \cite{bergamo2010exploiting,hua2015prajna,niu2016visual} have attracted more and more people's attention. Fergus et al. \cite{fergus2010learning} took the probabilistic latent semantic analysis (pLSA) technique to automatically select and learn object categories from web images. Hua et al. \cite{hua2015prajna} proposed the use of a clustering based method to filter ``group'' noisy images and a propagation-based method to filter individual noisy images. Niu et al.  \cite{niu2016visual} proposed to set each latent domain as a ``bag" and the images therein as ``instances", then image selection and classifier learning are formulated as a multi-instance learning problem. The advantage of these methods is that the need for manual intervention is eliminated. Unlike these studies, our proposed work simultaneously addresses the issues of image categorization and sub-categorization by levering the general corpus and web images. 

There are also few previous works \cite{mansur2008improving,hoai2013discriminative,ristin2015categories,wang2013max} dealing with the image sub-categorization problem. Mansur et al. \cite{mansur2008improving} proposed using probabilistic Latent Semantic Analysis (pLSA) to find subcategories and these subcategories are based on the similarity of images. In \cite{hoai2013discriminative}, both of positive and negative images are used to learn subcategories. Particularly, a new model by joint clustering and classification was proposed for discriminative sub-categorization. Wang et al. \cite{wang2013max} designed a formulation for dictionary learning (subcategories) by maximizing classification margins MIL. However, as some of the previous work \cite{bach2008diffrac}, methods \cite{mansur2008improving,hoai2013discriminative,wang2013max} still cannot assign semantic refinement labels for the newly discovered subcategories. Ristin et al. \cite{ristin2015categories} adopt the framework of Random Forests and proposed a regularized objective function that takes into account relations between categories and subcategories to improve the classification of subcategories. Unlike previous works, method \cite{ristin2015categories} can classify images into subcategories, but only NOUN subcategories. It is not comprehensive enough for describing the refinement images (e.g., category ``horse" in Fig. \ref{fig1}). In our work, we exploit general corpus information and web images for image categorization and sub-categorization. Our proposed approach can not only classify images into NOUN subcategories, but also into VERB, ADJECTIVE and ADVERB subcategories.

Motivated by the situation described above, we propose a novel automatically web-supervised image categorization and sub-categorization framework. In our work, we mainly consider the following two important issues: 1) the labels of subcategories derived from the general corpus usually have noise, how can we select useful labels of subcategories from these noisy labels; 2) the retrieved web images are often associated with inaccurate labels, so the learnt classifiers may be less robust, and the classification performance may be significantly degraded as well, how can we select useful images and learn domain robust classifiers from these noisy web training images. 

To find the labels of semantic refinement subcategories, we search the given categories in Google Books Ngram Corpus (GBNC) \cite{michel2011quantitative} with Parts-Of-Speech (POS), specifically with NOUN, VERB, ADJECTIVE and ADVERB. Further, as the labels of subcategories derived from the general corpus tend to have noise, we apply a relevance-based approach for removing noise and selecting the useful labels of subcategories. Finally, to cope with label noise of web training images, we treat each selected subcategory as a ``bag" and the images therein as ``instances". In specific, we propose a new multi-class MIL formulation to select the images from each bag and learn the classifiers for the categories and subcategories. Our aim is to select a subset of images from each bag to represent this bag, such that the training bags from all the categories can be well separated. To verify the superiority of our proposed approach, we conducted experiments on both image categorization and sub-categorization tasks. The experimental results demonstrated the superiority of our proposed approach. 

The main contributions of this work are summarized as follows:

\begin{enumerate}	
	\item Compared to existing methods, our proposed framework can not only classify images into NOUN subcategories, but also into VERB, ADVERB and ADJECTIVE subcategories. Our proposed framework has a better semantic refinement descriptions for the categories. 
	
	\item To suppress the search error and noisy subcategories (which are not filtered out) induced noisy images, we formulate image selection and classifier learning as a multi-class multi-instance learning problem and propose to solve the employed problem by the cutting-plane algorithm.
	
	\item We propose a new unified objective function to jointly learn the classifiers for categories and subcategories. Our proposed formulation can not only consider the relationship between category and its subcategories, but also consider the relationship between different categories. Thus, the classifiers in our work have a better domain adaptation ability.	
\end{enumerate} 

The rest of the paper is organized as follows. In Section \uppercase\expandafter{\romannumeral2}, a brief discussion of related works is given. Section \uppercase\expandafter{\romannumeral3} elaborates the proposed framework with the optimization algorithm. The experimental evaluations and discussions are presented in Section \uppercase\expandafter{\romannumeral4}. Lastly, the conclusion and future research directions are offered in Section \uppercase\expandafter{\romannumeral5}.

\section{Related work}

Due to the emergence of ImageNet, deep convolutional neural networks (CNN) have achieved a great success in image categorization. However, deep CNNs are computationally intensive and require a large number of labeled data. Simpler classifiers like support vector machine (SVM) \cite{weston1998multi} and nearest class mean classifiers (NCM) \cite{ristin2014incremental} provide us with another alternative which have much shorter running time and acceptable classification accuracy. 
To reduce the cost of manual annotation, some of the previous works also concentrated on the task of ``cleaning up" web images for training data collection. For example, Fergus et al. \cite{fergus2004visual} proposed the use of visual classifiers learned from Google image search engine to re-rank the images based on the visual consistency. Subsequent methods \cite{berg2006animals,li2010optimol,yao2016automatic} have employed similar removing mechanisms to automatically construct clean image datasets for training classifiers. In our work, we focus on another fundamental, yet often ignored, aspect of the problem: we argue that the current poor performance of classification models learned from the web is due to the selected images which may have different distributions with the test images.

Our work is also related to the recent works for latent domains discovering methods. In \cite{hoffman2012discovering}, Hoffman et al. proposed using a hierarchical clustering technique to find the feasible 
\begin{figure*}[tbp]
	\centering
	\includegraphics[width=0.98\textwidth]{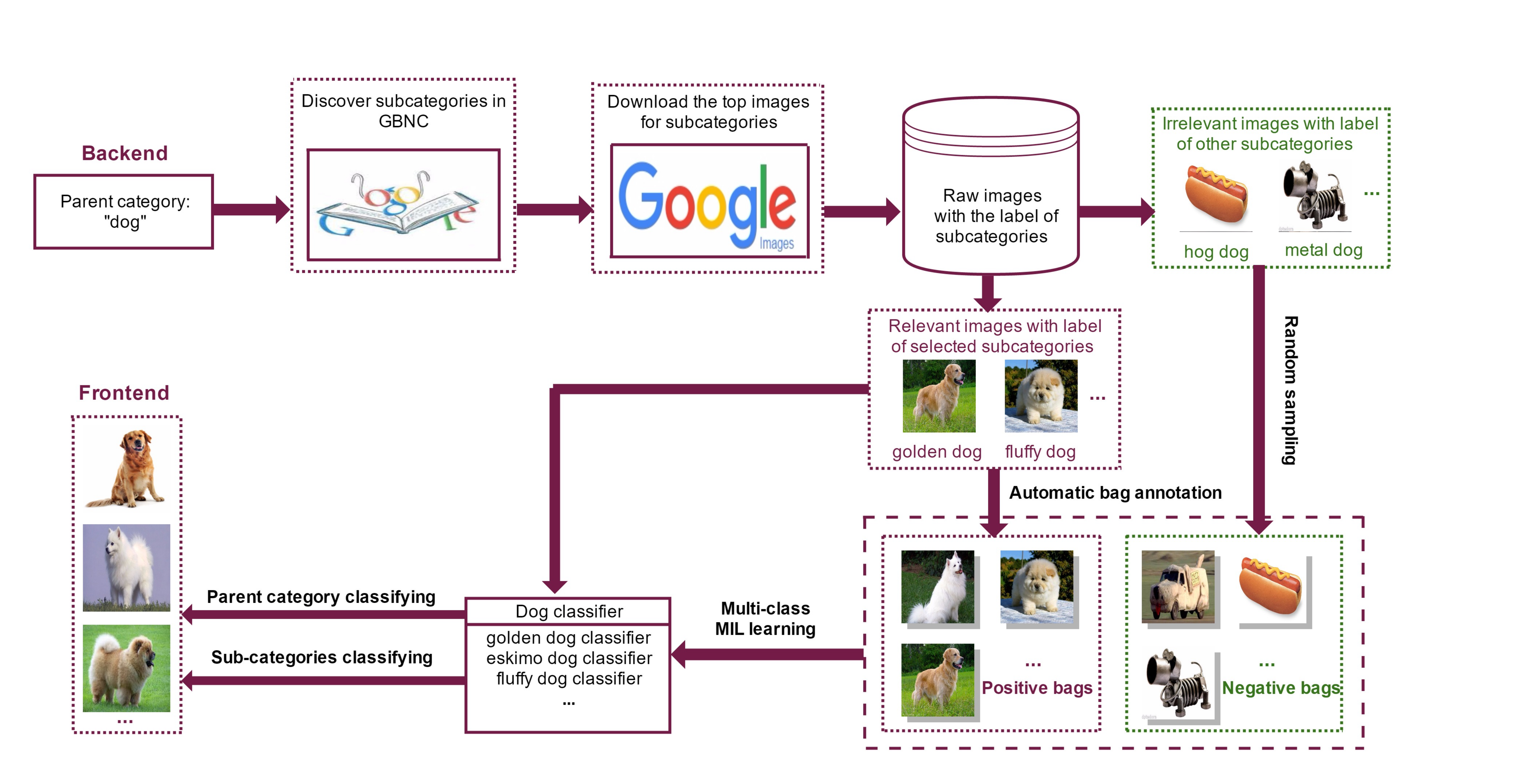}
	\caption{The proposed web-supervised image categorization and sub-categorization system. Backend and frontend work together as a pipeline for automatically collecting the labels of subcategories and associated images from the web, then selecting representative images and training classifiers for image categorization and sub-categorization.}
	\label{fig2}  
\end{figure*} 
latent domains while \cite{gong2013reshaping} adopt maximum distinctiveness and maximum learnability for different latent domains separation. Xiong et al. \cite{xiong2014latent} proposed a squared-loss mutual information based clustering model with category distribution priority in each domain to infer the domain assignment for images. The difference between our work and \cite{hoffman2012discovering,gong2013reshaping,xiong2014latent} is methods \cite{hoffman2012discovering,gong2013reshaping,xiong2014latent} cannot assign the semantic refinement labels to newly discovered latent domains. All the discovered latent domains still only have the label of the coarse category. In contrast, our work aims to classify images into categories and subcategories. All the images in our work will be assigned two labels including coarse category and refinement subcategory. So their motivations and formulations are inherently different from our work. 

Our work is more related to the recent image sub-categorization works \cite{hoai2013discriminative,ristin2015categories,wang2013max}, which assume several subcategories exist in each category. However, the subcategories discovered by \cite{hoai2013discriminative,wang2013max} still only have the label of the category. Refinement labels of subcategories are still unavailable. Ristin et al. \cite{ristin2015categories} adopt the framework of Random Forests and proposed a regularized objective function that takes into account relations between categories and subcategories to improve the classification accuracy. Unlike previous works, method \cite{ristin2015categories} can classify images into subcategories, but only the NOUN subcategories. The reason is that method \cite{ristin2015categories} relies on expert knowledge WordNet to obtain the semantic refinement subcategories. In our work, we eliminate the dependency on expert knowledge and propose to exploit general corpus information to obtain the semantic refinement subcategories. The advantage of our proposed approach is that our method can not only classify images into NOUN subcategories, but also into VERB, ADJECTIVE and ADVERB subcategories. 

As our approach relies on weakly labeled web images, it is loosely related to the multi-instance learning works \cite{andrews2002support,bunescu2007multiple,li2011text,li2009convex}. Method  \cite{bunescu2007multiple} and \cite{li2011text} proposed to partition the weakly labeled web images into a set of clusters and each cluster is treated as a ``bag", the images therein as ``instances". Correspondingly, different multi-instance learning methods were proposed in \cite{bunescu2007multiple} and \cite{li2011text}. In \cite{andrews2002support}, Andrews et al. adopt a heuristic way to iteratively train the image classifier and then infer the category labels of these instance images. Method \cite{li2009convex} proposed two convex optimization methods which maximize the margin of concepts via key instance generation at the instance-level and bag-level for locating the regions of interest (ROI) in the images automatically. 

Our work is largely inspired by the following work. A weakly supervised domain robust visual recognition system was recently proposed in \cite{niu2016visual} and achieved impressive performance for video event recognition. Niu et al. \cite{niu2016visual} first applied the latent domain discovering method \cite{gong2013reshaping} to find all the latent domains from the training data. Then the multi-instance learning method was leveraged to cope with noise in the labels of web training images. The main difference from ours is the formation process of the ``bags". Method \cite{niu2016visual} takes the latent domains as ``bags" while our method applies the selected subcategories. Compared to \cite{niu2016visual}, our selected ``bags" which derived from the general corpus have strong supervisory information. This supervisory information can help us to maximize the inter-class variation and simultaneously minimize the intra-class variation. In addition, as some of the previous works \cite{li2011text,li2009convex}, only category label can be assigned to images in \cite{niu2016visual} while our method can not only assign the category label, but also the refinement labels of subcategories. 

\section{Web-supervised image categorization and sub-categorization}

We seek to automate the process of classifying images into categories and subcategories by exploiting general corpus information and web images. As shown in Fig. \ref{fig2}, there is a backend subsystem (classifier building) and a frontend subsystem (categories and subcategories classifying). For the backend system, the input is a category label that we would like to build a classification model for. Then a set of semantically rich subcategories is obtained by searching in GBNC, from which the noisy subcategories are removed. After obtaining the candidate images by retrieving the labels of selected subcategories with image search engine, we treat each selected subcategory as a ``bag" and the images therein as ``instances". Particularly, we formulate image selecting and classifier learning as a multi-class MIL problem by selecting a subset of images from each bag to learn the classifiers. The final outputs of the backend system are (1) a classifier (e.g., ``dog") representing the category; (2) a set of classifiers (e.g, ``golden dog", ``fluffy dog" and ``Eskimo dog") corresponding to subcategories. The input of frontend is a set of images which will be given two labels including category and semantic refinement subcategories.

\subsection{Subcategories Discovering}

Inspired by recent works \cite{michel2011quantitative,yao2016automatic,yao2016domain}, we can use Google Books Ngram English 2012 Corpus to discover the labels of semantic refinement subcategories for modifying the given category. Our motivation is to find not only the semantically rich NOUN subcategories, but also VERB, ADJECTIVE and ADVERB subcategories. Compared to the expert knowledge WordNet and ConceptNet which only have NOUN subcategories, ngram data is much more general and exhaustive. Following \cite{lin2012syntactic} (see section 4.3), we specifically use the dependency gram data with parts-of-speech (POS) for refinement subcategories discovering. For example, given a parent category and its corresponding POS tag (e.g., `jumping, VERB'), we find all its occurrences annotated with POS tag within the dependency gram data. Of all the ngram dependencies retrieved for the given category, we choose those whose modifiers are tagged as NOUN, VERB, ADJECTIVE and ADVERB as the candidate subcategories. We utilize these semantic refinement subcategories (corresponding images) to reflect the different visual distributions of the category. The detailed subcategories discovered in this step can be found on website https://1drv.ms/f/s!Ahpq3qSTtg8NsxyjGslE2kjGcvTV. 

\begin{table}[thb]
	\centering
	\renewcommand{\arraystretch}{1.2}
	\caption{Examples of the candidate subcategories discovered by our approach.}
	\begin{tabular}{|c|l|}
		\hline
		\textbf{Category} &\textbf{ Discovered subcategories}  \\
		\hline
		\hline
		Horse & \tabincell{l}{\{jumping horse, grazing horse, rearing horse\}\\  \{plough horse, hunter horse, black horse\} \\ \{\textbf{wood horse, tang horse, betting horse, sea horse}\}}\\
		\hline	
		Boat  & \tabincell{l}{\{sails boat, fishing boat, diving boat\} \\ \{ski boat, tuna boat, bass boat\}\\ \{\textbf{leather boat, crystal boat, butter boat, paper boat}\}}\\
		\hline	
		Dog & \tabincell{l}{\{farm dog, wolf dog, fighting dog\}\\ \{pekingese dog, newfoundland dog, golden dog\}\\ \{\textbf{hot dog, cheese dog, metal dog, van dog}\}}\\
		\hline	
		Train & \tabincell{l}{\{subway train, metra train, electric train\}\\\{light train, double train, trolley train\} 	\\\{\textbf{potty train, storm train, column train}\}}\\		
		\hline
		Bird & \tabincell{l}{\{swallow bird, seagull bird, black bird\} \\ \{swan bird, eagle bird, humming bird\} 		\\\{\textbf{soup bird, angry bird, magic bird, bird nest}\}}  \\                        
		\hline
		Cat & \tabincell{l}{\{tiger cat, brown cat, hissing cat\} \\ \{fat cat, desert cat, ginger cat\} 		\\\{\textbf{lucky cat, tom cat, cat machine, missing cat}\}}  \\                        
		\hline				
	\end{tabular}
	\label{tab1}
\end{table}

\subsection{Noisy Subcategories Removing} 

Not all the discovered subcategories are useful, some noise may also be included (e.g., the bold subcategories in Table \ref{tab1}). Using these noisy subcategories to retrieve images for the category will have a negative effect on the accuracy and robustness of the classifier. To this end, we first remove these noisy subcategories before we select images and train classifiers for the category and subcategories. We retrieve the top $K$ images from image search engine for each candidate subcategory to represent their visual distributions. By analyzing the text semantics and visual distributions presented by these subcategories, we choose the following features to separate the useful subcategories from noise.

\subsubsection{Feature selecting}

From the visual relevance perspective, we want to eliminate visually less relevant subcategories (e.g., ``wood horse", ``paper boat"). The intuition is that relevant subcategories should have a relatively small inter-visual distance to its parent category and other relevant subcategories. We denote each image as $x_i$ and the compound feature $\phi_{k}=\frac{1}{k}\sum_{i=1}^{k}x_{i}$ of $K$ images in each subcategory to represent visual distribution of this subcategory. Suppose a parent category $C_i$ has $N$ subcategories, then we will have $N\times(N-1)/2$ inter-visual distances between subcategories. We calculate the minimum, maximum, average and standard deviation of the inter-visual distances between subcategories. Besides, we also calculate the inter-visual distance between subcategory and its parent category. Particularly, we denote these inter-visual distances by $ D=\{d_n, 0 \leqslant n<\frac{N\times (N-1)}{2}+N\} $ and normalize these distances to a number in [0,1] by:  
\begin{equation}{d_{n}}^{'} = \frac{d_{n}-\min\{d_{n}\}}{\max\{d_{n}\}-\min\{d_{n}\}}.\end{equation}

From the visual consistency perspective, we want to keep visually salient and eliminate non-salient subcategories (e.g., ``missing cat", ``betting horse"). The intuition is that visually salient subcategories have small intra-visual distance and exhibit predictable visual patterns. For the $K$ images in each subcategory, we calculate the $K\times(K-1)/2$ intra-visual distance among these $K$ images. We obtain the minimal, maximal, average and standard deviation of intra-visual distance like how we generate the inter-visual distance. Similarly, we normalize this intra-visual distances.

From the semantic relevance perspective, we want to remove semantically less relevant subcategories (e.g., ``tang horse", ``metal dog"). The intuition is that relevant subcategories tend to have a relatively small semantic distance to the parent category. Normalized Google Distance (NGD) constructs a method to extract semantic similarity distance from the World Wide Web (WWW) using Google page counts \cite{cilibrasi2007google}. For a search term $x$ and search term $y$, NGD is defined by:
\begin{equation}NGD(x,y)=\frac{\max\{\log f(x),\log f(y)\}-\log f(x,y)}{\log N-\min\{\log f(x),\log f(y)\}}\end{equation}
where $f(x)$ denotes the number of pages containing $x$, $f(x,y)$ denotes the number of pages containing both $x$ and $y$ and $N$ is the total number of web pages searched by Google.

In general, we derive the following features to separate the useful subcategories from noise:

$\bullet$ Normalized Google Distance between subcategory and its parent category

$\bullet$ Normalized visual distance between subcategory and its parent category

$\bullet$ Normalized minimum, maximum, average and standard deviation of the inter-visual distance between subcategory and other subcategories

$\bullet$ Normalized minimum, maximum, average and standard deviation of the intra-visual distance between different images in subcategory

\subsubsection{Classifier learning}

After deriving the selected features, we train a classifier to determine whether or not a subcategory should be selected. To train this classifier, we label a set of subcategories and this labeling work only needs to be done once for all categories. This classifier can be, for example, Support Vector Machine (SVM) based (which is also the one we used in our paper), Decision Tree based, etc. Although SVM is not the prevailing state-of-the-art method for classification, we find our approach to be effective in pruning noisy subcategories.

\subsection{Multi-class MIL Learning}

Although the Google image search engine has ranked the returned images, some noisy images may still be included. In addition, a few noisy subcategories which are not filtered out may also induce some noise. To this end, we propose our web-supervised multi-class MIL model for noisy images removing and domain robust classifiers learning.  

For ease of presentation, we denote each instance as $x_{i}$ with its label $y_{i}$ and each bag ${G_{m}}$ with the label ${Y_{m}}$. A matrix/vector is denoted by a uppercase/lowercase letter in boldface and the element-wise product between two matrices is represented by $\odot$. We also define the identity matrix as $\mathbf{I}$ and $\mathbf{0}$, $\mathbf{1}$ $\in \Re ^{n}$ denote the column vectors of all zeros and ones, respectively. The transpose of a vector or matrix is represented by $^\top$. The inequality $\textbf{u}=\left [u_{1},u_{2}...u_{n} \right ]^\top \geq \mathbf{0}$ means that $u_{i}\geq 0$ for \textit{i} = 1,...,\textit{n}. The indicator function is represented as $\delta\left ( i=j \right )$, where $\delta\left ( i=j \right )=1$ if $i=j$, and $\delta\left ( i=j \right )=0$, otherwise. 

\subsubsection{Formulation}

Since the retrieved web images may contain noise, we need to remove noise and select appropriate samples to train robust classifiers. To this end, a binary indicator $h_{i}\in \left \{ 0,1 \right \}$ is used to indicate whether or not training instance $x_{i}$ is selected. To be exact, $h_{i}=1$ when $x_{i}$ is selected, and $h_{i}=0$ otherwise. Since the precision of images returned from the Google image search engine tends to have a relatively high accuracy, we define each positive bag as at least having a portion of $\eta $ positive instances. The value of $\eta $ can be estimated from some prior knowledge \cite{li2011text,yao2016domain}. We define $\mathbf{h}=\left [ h_{1},...h_{N} \right ]^{\top}$ as the indicator vector, and use $ \mathrm{H} =\left \{ \mathbf{h} \right |\sum _{i\in I_{m}}h_{i}= \eta \left | G_{m} \right |,\forall m \}$ to represent the feasible set of $\mathbf{h}$, where $I_{m}$ represents the set of instance indices in $G_{m}$, and $\left | G_{m} \right |$ denotes the cardinality of $G_{m}$.
We assume there are $N$ retrieved web images coming from $C$ categories and $S$ subcategories. $z_{i,s}\in \left \{ 0,1 \right \}$ is a binary indicator variable and takes the value of 1 when $x_{i}$ belongs to the \emph{s}-th subcategory, and 0 otherwise. We denote $N_{s}=\sum _{i=1}^{N}z_{i,s}$ as the number of web training images from the \emph{s}-th subcategories. Based on MIL \cite{li2011text}, we propose our multi-class MIL formulation as follows:
\begin{equation}\label{eq3}
\begin{aligned}
& \underset{\mathbf{h},\mathbf{w}_{c,s},\xi_{m}}{\text{min}}
& & \frac{1}{2}\sum_{c=1}^{C}\sum_{s=1}^{S}\left \| \mathbf{w}_{c,s} \right \|^2+C_1\sum_{m=1}^{M}\xi_m  \\
\end{aligned}
\end{equation}
\begin{equation} \label{eq4}
	\begin{aligned}
		& & \mathrm{s.t.}\: \frac{1}{\left | G_m \right |}\sum _{i\in I_m}h_i\mathbf{(}\sum_{s=1}^{S}P_{i,s}{(\mathbf{w}_{Y_m,s})}^\top\phi (x_i)- \\
		& & {(\mathbf{w}_{\hat{c},\hat{s}})}^\top\phi (x_i)\mathbf{)}\geqslant \eta -\xi_m,\forall m,\hat{s},\hat{c}\neq Y_m \\
		& & \xi _m\geqslant 0, \forall m \\
	\end{aligned}
\end{equation}
where $C_1$ is a tradeoff parameter, $\xi _m$ are slack variables and $\phi (\cdot )$ is the feature mapping function. $P_{i,s}$ is the probability that the \emph{i}-th training sample comes from the \emph{s}-th subcategories. It can be obtained by calculate $P_{i,s}=(z_{i,s}/N_s)/\sum_{s=1}^{S}(z_{i,s}/N_s)$. The explanation for constraint \eqref{eq4} is that we force the total decision value (obtained by using the classifier for its own category) to be larger than those obtained by using the classifier for other categories. The motivation is we want to reduce the bag-level loss by removing the noise and identifying the good instances within training bags. 

\subsubsection{Solution}

Problem \eqref{eq3} is a non-convex mixed integer problem and is hard to solve directly. Inspired by recent works \cite{li2011text,li2009convex,kloft2011lp}, we can relax the dual form of \eqref{eq3} as a multiple kernel learning (MKL) problem which is much easier to solve. The derivations of \eqref{eq3} to its below dual form is provided in the Appendix A:
\begin{equation}\label{eq10}
\begin{aligned}
& & \min_\mathbf{h}\max_\mathbf{\bm{\alpha}} \: -\frac{1}{2}\bm{{\alpha}^{\top}}\textbf{Q}^{\mathbf{h}}\bm{\alpha}+\bm{\mathbf{{\zeta}^\top\mathbf{\alpha} }} \\
& & \mathrm{s.t.}\: \sum_{c,s}\alpha_{m,c,s}=C_1, \:\: \forall m, \\
& & \alpha _{m,c,s}\geqslant 0,\: \quad \forall m,c,s. \\
\end{aligned}
\end{equation}
$D=M\cdot C\cdot S$ and $\bm{\alpha}\in\mathbb{R}^{D}$ is a vector containing dual variables $\alpha _{m,c,s}$. $\bm{\zeta} \in \mathbb{R}^{D}$ is a vector, in which $\zeta _{m,c,s}=0$ if $c=Y_m$ and $\zeta _{m,c,s}=\eta $ otherwise. Each element in matrix $\textbf{Q}^{\mathbf{h}}\in \mathbb{R}^{D\times D}$ can be calculated through:
$\textbf{Q}^{\textbf{h}}= (1/\left|G_m\right|\left|G_{\hat{m}}\right|)\sum_{i\in I_m}\sum_{j\in I_{\hat{m}}}h_ih_j\o (\textbf{x}_i)^{\top}\o(\textbf{x}_j)\lambda (i,j,c,\hat{c},s,\hat{s})$.

Problem \eqref{eq10} is a mixed integer programming problem and is hard to directly optimize the indicator vector \textbf{h}. Inspired by recent work \cite{niu2016visual}, we can find the coefficients of
$\mathbf{h}_{t}{\mathbf{h}}^{\top}_{t}$. For consistent presentation, we denote $\mathbf{d}=[d_1,...d_T]^{\top}$, $T = \left |\mathrm{H}  \right |$ and the feasible set of $\bm{\alpha}$ and \textbf{d} as $\nu $ and $D=\{\mathbf{d}|\mathbf{d^{\top}}\bm{1}=1,\mathbf{d}\geqslant 0 \}$, respectively. Then we can get the following optimization problem:
\begin{equation}\label{eq11}
\begin{aligned}
& \min_{\mathbf{d}\in D}\max_{\bm{\alpha} \in \nu }
& & -\frac{1}{2}\sum_{t=1}^{T}d_{t}\bm{\alpha}^{\top}\textbf{Q}^{\bm{\mathrm{h}}_t}\bm{\alpha}+\bm{\zeta} ^{\top}\bm{\alpha}. \\
\end{aligned}
\end{equation}
When we set the base kernel as $\textbf{Q}^{\bm{\mathrm{h}}_t}$, the above problem is similar to the MKL dual form and we are able to solve it on its primal form, which is a convex optimization problem:
\begin{equation}\label{eq12}
\begin{aligned}
\min_\mathbf{\mathbf{d}\in D,\bm{\mathrm{w}}_{t},\xi_{m}}\: \frac{1}{2}\sum_{t=1}^{T}\frac{\left \| \mathbf{\mathrm{w}}_{t} \right \|^2}{d_t}+C_1\sum_{m=1}^{M}\xi_m   \\
\end{aligned}
\end{equation}
\begin{equation}\label{eq13}
\begin{aligned}
\mathrm{s.t.}\: \sum_{t=1}^{T}\mathrm{\mathbf{w}}_t^{\top}\varphi (\mathrm{\mathbf{h}}_t,G_m,c,s)\geqslant \zeta _{m,c,s}-\xi _m,\: \forall m,c,s \\
\end{aligned}
\end{equation}
where $\varphi(\bm{\mathrm{h}}_t,G_m,c,s)$ is the feature mapping function induced by $\textbf{Q}^{\mathbf{h}_t}$. The derivations of \eqref{eq11} is the dual form of \eqref{eq12} are provided in the Appendix B. In the next, we will give a solution to \eqref{eq12}.

We solve the convex problem in \eqref{eq12} by updating $\mathbf{d}$ and $\{ \mathbf{w_t},\xi _m \}$ in an alternative way.

$\bullet$ $\emph{Update}$ \textbf{d}: We firstly fix $\{ \mathbf{w_t},\xi _m \}$ to solve \textbf{d}. By introducing a dual variable $\beta $ for constraint $\mathbf{d^{\top}}\bm{1}=1$, the Lagrangian form of \eqref{eq12} can be derived as:   
\begin{equation}\label{eq16}
\begin{aligned}
\pounds =\frac{1}{2}\sum_{t=1}^{T}\frac{\left \| \mathbf{\mathrm{w}}_{t} \right \|^2}{d_t}+C_1\sum_{m=1}^{M}\xi_m -\sum _{m,c,s}\alpha_{m,c,s}  \\
(\sum_{t=1}^{T}\mathrm{\mathbf{w}}_t^{\top}\varphi (\mathrm{\mathbf{h}}_t,G_m,c,s)- \zeta _{m,c,s}+\xi _m)+\beta(\sum_{t=1}^{T}d_t-1).
\end{aligned}
\end{equation}

Through set the derivative of \eqref{eq16} with respect to $d_t$ as zero, we can get:
\begin{equation}\label{eq18}
\begin{aligned}
d_t= \frac{\left \| \mathbf{w}_{t} \right \|}{\sqrt{2\beta }},\: \forall t=1,...,T.
\end{aligned}
\end{equation}
For parameter $\beta$, ${\left \| \mathbf{w}_{t} \right \|}/{\sqrt{2\beta }}$ is monotonically decreasing. In addition, parameter $d_{t}$ satisfy $\sum_{t=1}^{T}d_t=1$. Therefore, we can use binary search method to solve $\beta$ and recover $d_{t}$ according to \eqref{eq18}.   

$\bullet$ $\emph{Update}$ $\bm{\mathrm{w}}_t$: When $\bm{\mathrm{d}}$ is fixed, $\bm{\mathrm{w}}_t$ can be obtained by solving $\bm{\mathrm{\alpha }}$ in \eqref{eq11}. Problem \eqref{eq11} is a quadratic programming problem w.r.t $\bm{\mathrm{\alpha }}$. Since there are $M\cdot C\cdot S$ variables in our problem, it is time-consuming to employ the existing quadratic programming solvers. Inspired by recent works \cite{gehler2008infinite,li2009tighter}, we apply the cutting-plane algorithm \cite{kelley1960cutting} to solve this quadratic programming problem.  

We start from a small number of base kernels and at each iteration we add a new violating base kernel. Therefore, only a small set of $\mathbf{h}$ need to be solved at each iteration and the whole problem can be optimized more effectively. By setting the derivatives of \eqref{eq16} with respect to $\{\mathbf{w}_t,\xi_t,d_t\}$ as zeros, \eqref{eq11} can be rewritten as:
\begin{equation}\label{eq19}
\begin{aligned}
&	\max_{\beta ,\bm{\alpha}\in \nu}-\beta +\bm{\zeta} ^{\top}\bm{\alpha } \\
&	\mathrm{s.t.}\: \frac{1}{2}\bm{\alpha}^{\top}\textbf{Q}^{\mathbf{h}_t}\bm{\alpha} \leqslant \beta , \: \forall t. 
\end{aligned}
\end{equation}

We solve \eqref{eq19} by solving $\alpha$ with only one constraint at the first, then add a new violating constraint iteratively. Particularly, since each constraint is associated with an $\mathbf{h}_t$ , we can obtain the most violated constraint by optimizing:
\begin{equation}\label{eq20}
\begin{aligned}
& & \max_\mathbf{h}\: \frac{1}{2}\bm{\alpha}^{\top}\textbf{Q}^{\mathbf{h}}\bm{\alpha}  \\
\end{aligned}
\end{equation} 
After a simple derivation, we can rewrite \eqref{eq20} as:
\begin{equation}\label{eq21}
\begin{aligned}
\max_\mathbf{h}\: \mathbf{h}^{\top}(\frac{1}{2}\hat{\textbf{Q}}\odot  (\hat{\bm{\alpha}}\hat{\bm{\alpha}}^{\top}  ))\mathbf{h}  \\
\end{aligned}
\end{equation}
where $\hat{\alpha }_i=1/\left | G_m \right |\sum _{c,s}\alpha _{m,c,s}$ for $i\in I_m$ and $\hat{\textbf{Q}}=\sum _{c,\hat{c},s,\hat{s}}\phi (x_i)^{\top}\phi (x_j)\lambda (i,j,c,\hat{c},s,\hat{s})$. Problem \eqref{eq21} can be solved approximately through enumerate the binary indicator vector $\mathbf{h}$ in 
\begin{algorithm}[tb]
	\caption{Cutting-plane algorithm for solving the proposed web-supervised multi-class MIL model.}
	\begin{algorithmic}[1]
		\REQUIRE ~~\\
		Auto-labelled image bags $\{(G_m,Y_m)|_{m=1}^M\}$, initialize $y_{i} = 1$ for all $ x_{i} $ in selected bags $ G_m $.\\
		\STATE Set $t=1$ and $C=\{\mathbf{h}_1\}$;   \\
		\STATE \textbf{Repeat}\\
		\STATE \quad $t =  t + 1$; \\
		\STATE \quad Compute MKL to solve $\textbf{d}$ and \bm{$ \alpha $} in \eqref{eq11} based on \textit{C}; \\
		\STATE \quad //\textit{Find the most violating $\mathbf{h}_t$}   \\
		\STATE \quad \textbf{for} each bag $ G_{m} $
		\STATE \quad \quad  Fix the labelling of instances in all other bags;
		\STATE \quad \quad  Enumerate the candidates of $ y_{i} $ in $ G_{m} $;     \\
		\STATE \quad \quad  Find the optimal $ \mathrm{\mathbf{y}}_m $ by maximizing \eqref{eq21};  \\
		\STATE \quad \textbf{end}
		\STATE \quad $\mathbf{repeat}$ lines 6-10 $\mathbf{until}$ there is no change in \textbf{h};		
		\STATE \quad Add the most violating $\mathbf{h}_t$ to the violation set $C = $   \\
		\STATE \quad $C \cup \mathbf{h}_t$; \\
		\STATE \textbf{Until} The objective of \eqref{eq11} converges.\\  
		\ENSURE ~~\\		
		The learnt image classifier \emph{f}(x). 		
	\end{algorithmic}
	\label{alg}
\end{algorithm}
a bag by bag fashion iteratively to maximize \eqref{eq21} until there is no change in $\mathbf{h}$. The detailed solutions of cut-plane algorithm for our web-supervised multi-class multi-instance learning model are described in Algorithm 1.

Since the visual distributions of the training samples from same category or subcategory are generally more similar than different categories and subcategories, we train one classifier for each category and each subcategory. In general, a total of $C\times S$ classifiers $f_{c,s}(x)|c=1,...C,s=1,...S$ will be learned. $f_{c,s}(x)={(\textbf{w}_{c,s})}^{\top}\o (x)$ (for better representation, we omit the bias term here) represents the classifier of the \emph{s}-th subcategory and the \emph{c}-th category. The decision function for category $\emph{C}$ is obtained by integrating the learned classifiers from multiple subcategories: $f_c(x_i)=\sum_{s=1}^{S}P_{i,s}f_{c,s}(x_i)$.   

Given a testing image \textbf{x}, we want to find the labels of the most matched subcategory and category, whose classifier achieves the largest decision value from all the subcategories and categories respectively. Thus, the subcategory label of image \textbf{x} can be predicted by:
\begin{equation}
\begin{aligned}
\arg \max_s {\textbf{w}^{\top}_{c,s}}\phi (\textbf{x})
\end{aligned}
\end{equation}
and the category label by:
\begin{equation}
\begin{aligned}
\arg \max_{c} (\max_s {\textbf{w}^{\top}_{c,s}}\phi (\textbf{x})).
\end{aligned}
\end{equation}

In summary, to suppress the search error induced noisy images, we propose a multi-class MIL model to select a subset of training images from selected bags and simultaneously learning the optimal classifiers based on this selected images. 

\section{Experiments}

In this section, we first conduct experiments on both image categorization and sub-categorization to demonstrate the superiority of our proposed approach. Then we quantitative analyse the role of different steps contributing to the final results. In addition, we also analyse the parameter sensitivity and time complexity of our proposed approach in this section.

\subsection{Image Categorization}

The goal of this experiment is to compare the image categorization ability of our proposed approach with other related works.

\subsubsection{Experimental setting}
We follow the setting in \cite{bergamo2010exploiting} and exploit web images as
the training set, human-labelled images as the testing set. Particularly, we evaluate the performance on the following dataset:

$\bullet$ PASCAL VOC 2007 \cite{everingham2010pascal}. The PASCAL VOC 2007 dataset contains 9963 images in 20 categories. Each category has training/validation data and test data. For this experiment, we only use the test data in PASCAL VOC 2007 as the benchmark testing set. The detailed number of images for each category in this experiment is shown in Table \ref{tab_new1}.

$\bullet$ STL-10 \cite{coates2010analysis}. The STL-10 dataset has ten categories, and each category of which contains 500 training images and 800 test images. All of the images in STL-10 are color 96 $\times$ 96 pixels. We also use the test images in STL-10 as the benchmark testing set. 

$\bullet$ CIFAR-10 \cite{krizhevsky2009learning}. The CIFAR-10 dataset consists of 60000 32$\times$32 images in 10 categories, with 6000 images per category, of which 5000 are training images and 1000 are test images. Similarly, we only use the test images in CIFAR-10 as the benchmark testing set.

For each category, we first obtain the semantic refinement subcategories by searching in the Google Books Ngram Corpus. Then we retrieve the top 100 images from the Google image search engine for each candidate subcategory to represent its visual distribution. Particularly, we have released the discovered candidate subcategories and retrieved images for all the candidate subcategories on website https://1drv.ms/f/s!Ahpq3qSTtg8NsxyjGslE2kjGcvTV. 

We calculate the Normalized Google Distance between subcategory and its parent category. We obtain the center of each subcategory by using the K-means clustering algorithm \cite{pelleg2000x} (k=1), then we calculate the normalized visual distance (Euclidean distance) between subcategory and its parent category; calculate the normalized minimum, maximum, average and standard deviation of inter visual distance between subcategory and other subcategories; calculate the normalized minimum, maximum, average and standard deviation of intra-visual distance between different images in subcategory. We label a set of 500 positive samples and 500 negative samples to learn the linear SVM classifier for removing noisy subcategories and selecting useful subcategories.

After we obtain the selected subcategories, the first $\bold{M}=50$ images were selected for constructing the positive bags 
\begin{table}[tb]
	\centering
	\renewcommand{\arraystretch}{1.2}
	\caption{The detailed number of images used for categorization in PASCAL VOC 2007.}
	\begin{tabular}{|p{1.5cm}<{\centering}|p{1.5cm}<{\centering}|p{1.5cm}<{\centering}|p{1.5cm}<{\centering}|}
		\hline		
		\textbf{Category} & \textbf{Number} & \textbf{Category} &\textbf{ Number}    \\
		\hline	
		Aero     & 204 & tabl  & 190   \\		
		\hline	 
		Bicycle  & 239 & Dog   & 418     \\
		\hline	
		Bird     & 282 & Horse & 274     \\
		\hline	
		Boat     & 172 & Mbike & 222    \\
		\hline
		Bottle   & 212 & Person& 2007     \\
		\hline	
		Bus      & 174 & Plant & 224     \\
		\hline
		Car      & 721 & Sheep & 97     \\
		\hline	
		Cat      & 322 & Sofa  & 223     \\
		\hline		
		Chair    & 417 & Train & 259    \\
		\hline
		Cow      & 127 & Tv    & 229    \\
		\hline				    	    	    	    	    	    	     	       						
	\end{tabular}
	\label{tab_new1}
\end{table} 
which corresponding to the selected subcategories. Negative bags can be obtained by randomly sampling a few irrelevant images. By treating each selected subcategory as a ``bag" and the images therein as ``instances", we formulate a multi-class MIL method to select a subset of training images from each bag and simultaneously learning the optimal classifiers based on the selected images. Particularly, we define each positive bag as having at least a portion of $\eta  =0.7 $ positive instances and set the tradeoff parameter $ C_{1} = 10^{-1} $. We will discuss the parameter setting more details in Section IV-D. 
To compare with other baseline methods, we evenly select 500 images from positive bags for each category as the training set. For this experiment, the feature is dense HOG \cite{dalal2005histograms}.

\subsubsection{Baselines} In order to quantify the performance of our proposed approach, four set of weakly supervised or web-supervised baselines are selected to compare with our proposed approach: 

$\bullet$ SVM method. The SVM method includes multi-class SVM \cite{weston1998multi}. For the multi-class SVM, the 500 training images for each category are directly retrieved from the image search engine with the category label.  

$\bullet$ MIL methods. The MIL methods contain instance level method mi-SVM \cite{andrews2002support} and bag level method sMIL \cite{bunescu2007multiple}. For method mi-SVM, the training images are also retrieved from the image search engine. Particularly, we take the proposed heuristic way to iteratively select 500 images for each category and train the image classifier. For method sMIL, we firstly retrieve the candidate images from the image search engine, then we partition the candidate images into a set of clusters. Each cluster is treated as a ``bag" and the images therein as ``instances". Correspondingly, we take the proposed MIL method to select the 500 training images for each category and train the image classifier.

\begin{figure*}[tbp]
	\centering
	\includegraphics[width=0.97\textwidth]{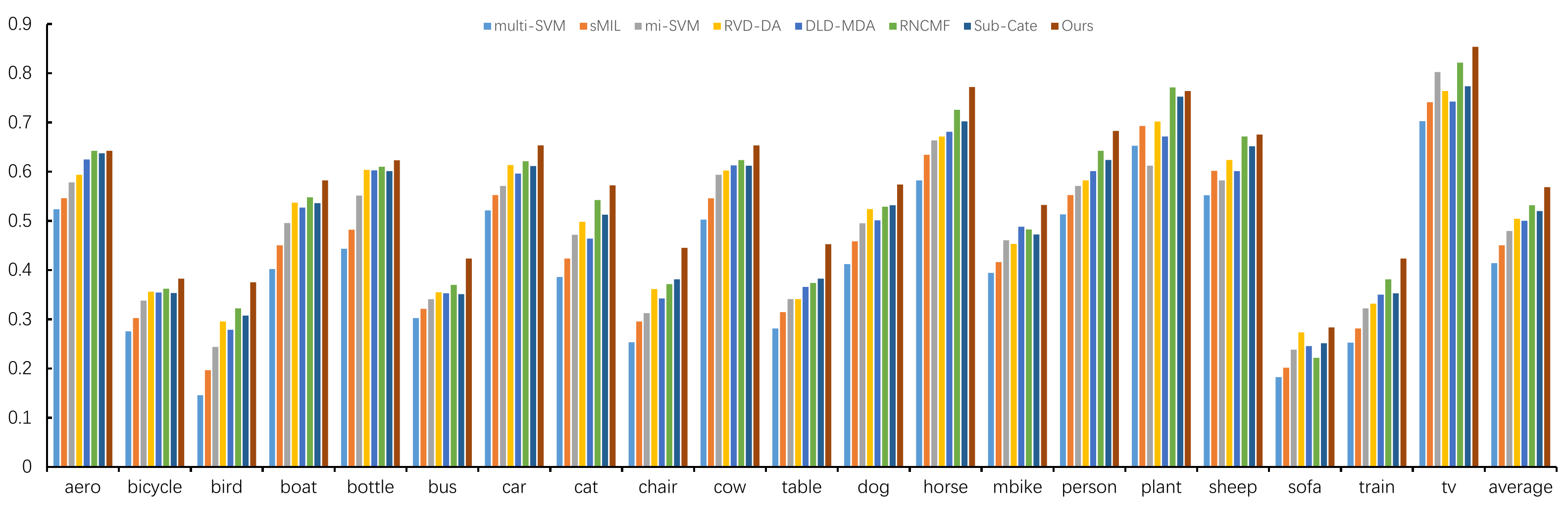}
	\caption{The detailed performance comparison of classification accuracy over 20 categories on the PASCAL VOC 2007 dataset.}
	\label{fig3} 
\end{figure*} 

\begin{table*}[htb]
	\centering
	\renewcommand{\arraystretch}{1.2}
	\caption{The detailed performance comparison of classification accuracy over 10 categories on the STL-10 dataset.}
	\begin{tabular}{|p{2cm}<{\centering}|p{0.95cm}<{\centering}|p{0.95cm}<{\centering}|p{0.95cm}<{\centering}|p{0.95cm}<{\centering}|p{0.95cm}<{\centering}|p{0.95cm}<{\centering}|p{0.95cm}<{\centering}|p{0.95cm}<{\centering}|p{0.95cm}<{\centering}|p{0.95cm}<{\centering}|p{0.95cm}<{\centering}|}
		\hline
		\multirow{2}{*}{Method} & \multicolumn{10}{c|}{Category} & \multirow{2}{*}{Average} \\
		\cline{2-11}
		
		& airplane &  bird  &  car  &  cat  &  deer  &  dog  &  horse  &  monkey  &  ship  &  truck  &     \\
		\hline
		\hline	 
		multi-SVM   & 0.497 & 0.193 & 0.384 & 0.222 & 0.467 & 0.293 & 0.534 & 0.355 & 0.443 & 0.348 & 0.374   \\
		\hline	
		sMIL        & 0.526 & 0.221 & 0.433 & 0.285 & 0.504 & 0.339 & 0.593 & 0.427 & 0.523 & 0.381 & 0.423  \\
		\hline	
		mi-SVM      & 0.531 & 0.242 & 0.464 & 0.283 & 0.528 & 0.346 & \textbf{0.642} & 0.404 & 0.547 & \textbf{0.482} & 0.444  \\
		\hline
		DLD-MDA     & 0.549 & 0.265 & 0.483 & 0.335 & 0.542 & 0.363 & 0.613 & 0.477 & 0.556 & 0.434 & 0.461   \\
		\hline	
		RVD-DA      & 0.557 & 0.271 & 0.488 & 0.326 & 0.547 & 0.352 & 0.608 & 0.484 & 0.567 & 0.446 & 0.458   \\
		\hline
		Sub-Cate    & 0.553 & \textbf{0.294} & 0.482 & 0.331 & 0.535 & 0.354 & 0.616 & 0.456 & 0.550 & 0.443 & 0.465   \\
		\hline	
		RNCMF       & 0.573 & 0.271 & 0.486 & 0.336 & 0.562 & 0.371 & 0.613 & 0.463 & 0.553 & 0.446 & 0.467   \\
		\hline		
		Ours        & \textbf{0.596} & 0.284 & \textbf{0.516} & \textbf{0.366} & \textbf{0.582} & \textbf{0.397} & 0.636 & \textbf{0.502} & \textbf{0.582} & 0.473 & \textbf{0.493}  \\
		\hline				    	    	    	    	    	    	     	       						
	\end{tabular}
	\label{tab2}
\end{table*}

\begin{table*}[htb]
	\centering
	\renewcommand{\arraystretch}{1.2}
	\caption{The detailed performance comparison of classification accuracy over 10 categories on the CIFAR-10 dataset.}
	\begin{tabular}{|p{2cm}<{\centering}|p{0.95cm}<{\centering}|p{0.95cm}<{\centering}|p{0.95cm}<{\centering}|p{0.95cm}<{\centering}|p{0.95cm}<{\centering}|p{0.95cm}<{\centering}|p{0.95cm}<{\centering}|p{0.95cm}<{\centering}|p{0.95cm}<{\centering}|p{0.95cm}<{\centering}|p{0.95cm}<{\centering}|}
		\hline
		\multirow{2}{*}{Method} & \multicolumn{10}{c|}{Category} & \multirow{2}{*}{Average} \\
		\cline{2-11}
		
		& airplane &  car  &  bird  &  cat  &  deer  &  dog  &  frog  &  horse  &  ship  &  truck  &     \\
		\hline
		\hline	 
		multi-SVM   & 0.397 & 0.304 & 0.094 & 0.163 & 0.345 & 0.264 & 0.153 & 0.423 & 0.342 & 0.277 & 0.276   \\
		\hline	
		sMIL        & 0.423 & 0.302 & 0.073 & 0.241 & 0.352 & 0.285 & 0.201 & 0.453 & 0.325 & 0.271 & 0.293  \\
		\hline	
		mi-SVM      & 0.422 & 0.328 & 0.103 & 0.232 & 0.354 & 0.323 & 0.203 & 0.414 & 0.372 & 0.274 & 0.302  \\
		\hline
		DLD-MDA     & 0.451 & 0.293 & 0.124 & 0.271 & 0.322 & 0.335 & 0.197 & 0.393 & 0.337 & 0.265 & 0.298   \\
		\hline	
		RVD-DA      & \textbf{0.482} & 0.322 & 0.124 & 0.284 & 0.353 & 0.343 & 0.204 & 0.424 & 0.363 & 0.285 & 0.314   \\
		\hline
		Sub-Cate    & 0.433 & 0.303 & \textbf{0.132} & 0.283 & 0.363 & \textbf{0.352} & 0.214 & 0.454 & 0.323 & 0.285 & 0.318   \\
		\hline	
		RNCMF       & 0.411 & 0.315 & 0.124 & 0.314 & 0.444 & 0.343 & 0.208 & 0.464 & 0.342 & 0.295 & 0.326   \\
		\hline
		Ours        & 0.452 & \textbf{0.373} & 0.113 & \textbf{0.342} & \textbf{0.493} & 0.322 & \textbf{0.265} & \textbf{0.502} & \textbf{0.382} & \textbf{0.322} & \textbf{0.357}  \\
		\hline				    	    	    	    	    	    	     	       						
	\end{tabular}
	\label{tab3}
\end{table*}

$\bullet$ Latent domain discovering methods. The latent domain discovering methods include two methods DLD-MDA \cite{hoffman2012discovering} and RVD-DA\cite{gong2013reshaping}. For method DLD-MDA, we firstly obtain the candidate images from the image search engine, then we take the hierarchical clustering technique to find the feasible latent domains. By treating each latent domain as a ``bag" and the images therein as ``instances", we take 
the proposed MIL method to select 500 images for each category and train the image classifier. For method RVD-DA, after we obtain the candidate images from the image search engine, we take the proposed maximum distinctiveness and maximum learnability to find and separate the latent domains. Similarly, we take the proposed MIL method to select 500 images for each category and train the image classifier.

$\bullet$ Sub-categorization methods. The sub-categorization methods Sub-Cate\cite{hoai2013discriminative} and RNCMF\cite{ristin2015categories} are also used to do image categorization. For method Sub-Cate, the candidate images are retrieved from the image search engine. Then we discover the subcategories during these candidate images by joint clustering and classification. We evenly select 500 images from these subcategories and train image classifiers. For method RNCMF, we also obtain the candidate images from the image search engine. We take the framework of Random Forests and the proposed regularized objective function to select 500 images for each category and train the image classier.  

For all the baseline methods, there are some parameters to be set in advance. All the training images for each category are obtained by retrieving from the Google image search engine. For the other parameters, we adopt the same parameter configuration as described in their original works.

\subsubsection{Experimental results}

The experimental results are summarized in Fig. \ref{fig3}, Table \ref{tab2} and Table \ref{tab3}. From the results, we make the following observations:

During the 20 categories in PASCAL VOC 2007, we achieved the best results in 19 categories. In the 10 categories of STL-10 and CIFAR-10, we obtained the best results in 7 categories respectively. In addition, our approach also achieved the best average results on all three datasets.

\begin{figure}[tbp]
	\centering
	\includegraphics[width=0.42\textwidth]{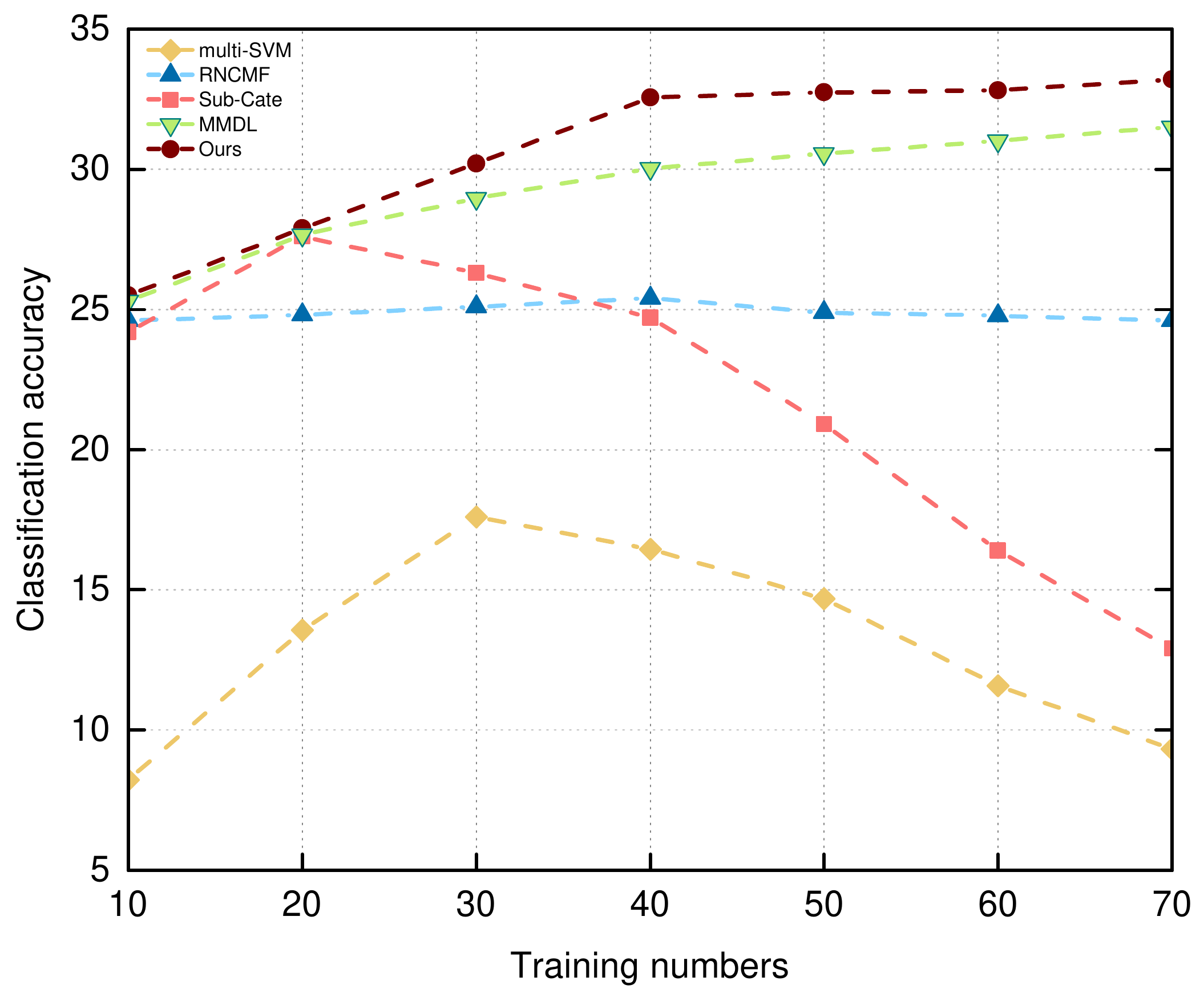}
	\caption{Sub-categorization accuracy (\%) of the different methods using 50 testing images and a varying number of web training images for per subcategory.}
	\label{fig4} 
\end{figure}    

We observe that the MIL learning methods \cite{bunescu2007multiple,andrews2002support}, the latent domain discovering methods \cite{hoffman2012discovering,gong2013reshaping}, the sub-categorization methods \cite{hoai2013discriminative,ristin2015categories} and our method are generally better than SVM method \cite{weston1998multi}. One possible explanation is that additional information like ``bags", latent domains or subcategories are beneficial in web training images for image categorization. 
In specific, MIL learning methods sMIL and mi-SVM achieve better results than SVM method multi-SVM on all three datasets. The explanation is perhaps that it is necessary to remove noisy images from the training set during the process of classifier learning. Learning directly from the web images without noise removing may affect the performance of the classifier due to the presence of noisy images in the training data. 

Sub-categorization methods RNCMF and Sub-Cate generally perform better than other two latent domains discovering methods DLD-MDA, RVD-DA and other MIL, SVM baseline methods. One possible explanation is that classifiers learned from Sub-categorization methods which exploiting subcategories to learn integrated classifiers are more domain robust than MIL methods, domain discovering methods and SVM baseline methods for image categorization. 

It is interesting to observe that all classifiers have a relatively poor performance on dataset CIFAR-10 and STL-10 than on dataset PASCAL VOC 2007. The explanation is perhaps that all images in CIFAR-10 are cut to 32$\times$32 and in STL-10 are 96$\times$96. Objects in these small images are placed in the middle of the image. However, our web training images and the testing images in PASCAL VOC 2007 are both full size and contain relatively more additional objects or scenes in images. 

Finally, our proposed approach achieves the best average performance on all three datasets, which demonstrate the superiority of our approach. The reason is our method simultaneously uses the MIL technique for handling label noise in the web training images and exploits multiple subcategories to learn integrated classifiers. Compared to SVM method, our method not only removes the noise, but also utilizes subcategories to learn integrated domain robust classifiers. Compared to MIL, latent domain discovering and sub-categorization 
\begin{figure}[tbp]
	\centering
	\includegraphics[width=0.42\textwidth]{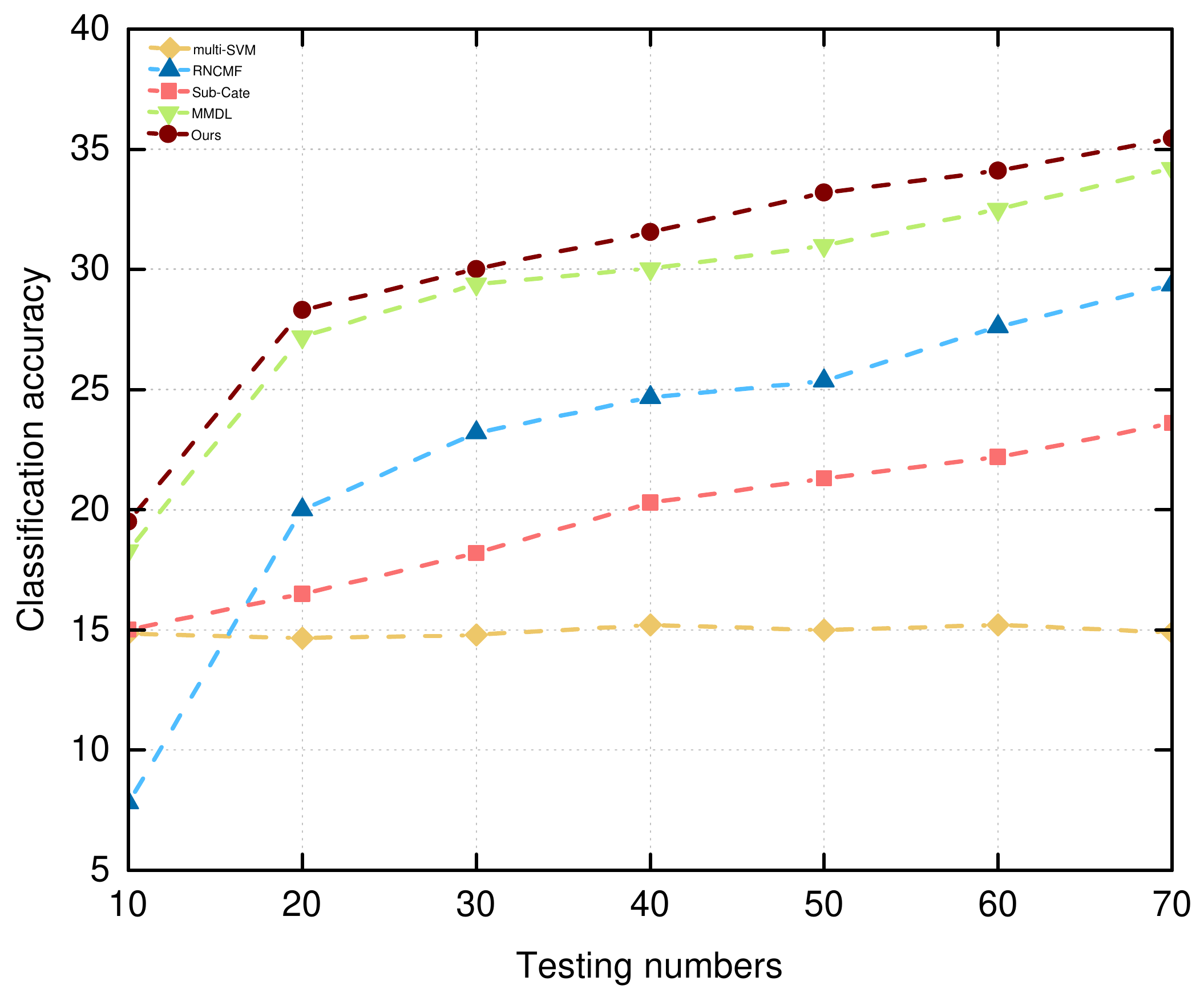}
	\caption{Sub-categorization accuracy (\%) of the different methods using 50 web training images and a varying number of testing images for per subcategory.}
	\label{fig5}
\end{figure}
methods, the multiple subcategories in our method have strong supervisory information which was obtained from the perspective of text semantics (e.g., subcategories discovering and noisy subcategories removing). This strong supervisory information can help us to maximize the inter-class variation and simultaneously minimize the intra-class variation.

\subsection{Image Sub-categorization} 

The objective of this experiment is to compare the image sub-categorization ability of our method with four weakly supervised or web-supervised baseline methods.

\subsubsection{Experimental setting}

For image sub-categorization, we choose a subset of ImageNet as the benchmark dataset for testing different methods. The reason is that ImageNet which constructed according to the WordNet has a hierarchy structure. In particular, we select five categories including ``airplane", ``bird", ``cat", ``dog" and ``horse" as the parent categories and all their leaf synsets as the subcategories. We are only concerned with the two-tier structure and deeper structure synsets are ignored. Thus, we obtain 5 parent categories and 97 subcategories. A detailed number of subcategories and corresponding images for each category in ImageNet for this experiment can be found in Table \eqref{tab_new2}. We retrieve the top 100 images for each subcategory from the image search engine as the common original training images. So we have a total of 9700 training images for 5 parent categories and 97 subcategories. For a fair comparison with other baseline methods, we replace the subcategories discovering and noisy subcategories removing procedures with the given parent categories and subcategories from ImageNet in our work. So the initial value of $C$ and $M$ in our work is 5 and 97 respectively. For this experiment, the feature is 1000-dimensional bag-of-visual-words (BoW) based on densely sampled SIFT features \cite{lindeberg2012scale}. The detailed list of the 97 subcategories in ImageNet and the common original 9700 web training images can be found on website https://1drv.ms/f/s!Ahpq3qSTtg8NsxyjGslE2kjGcvTV.

\begin{figure*} [t]
	\centering
	\subfloat[]{%
		\includegraphics[width=1.7in, height=1.7in]{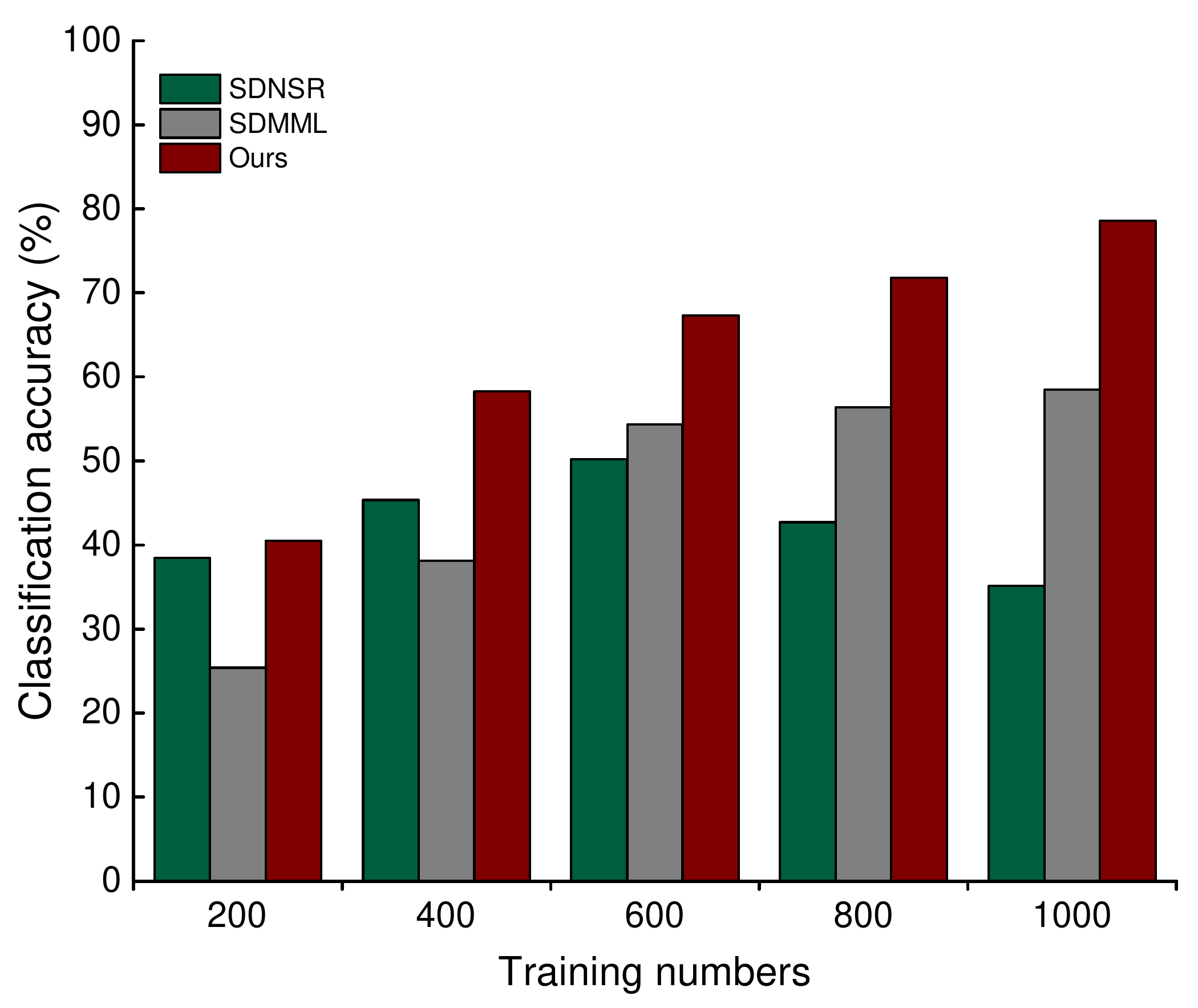}}
	\subfloat[]{%
		\includegraphics[width=1.7in, height=1.7in]{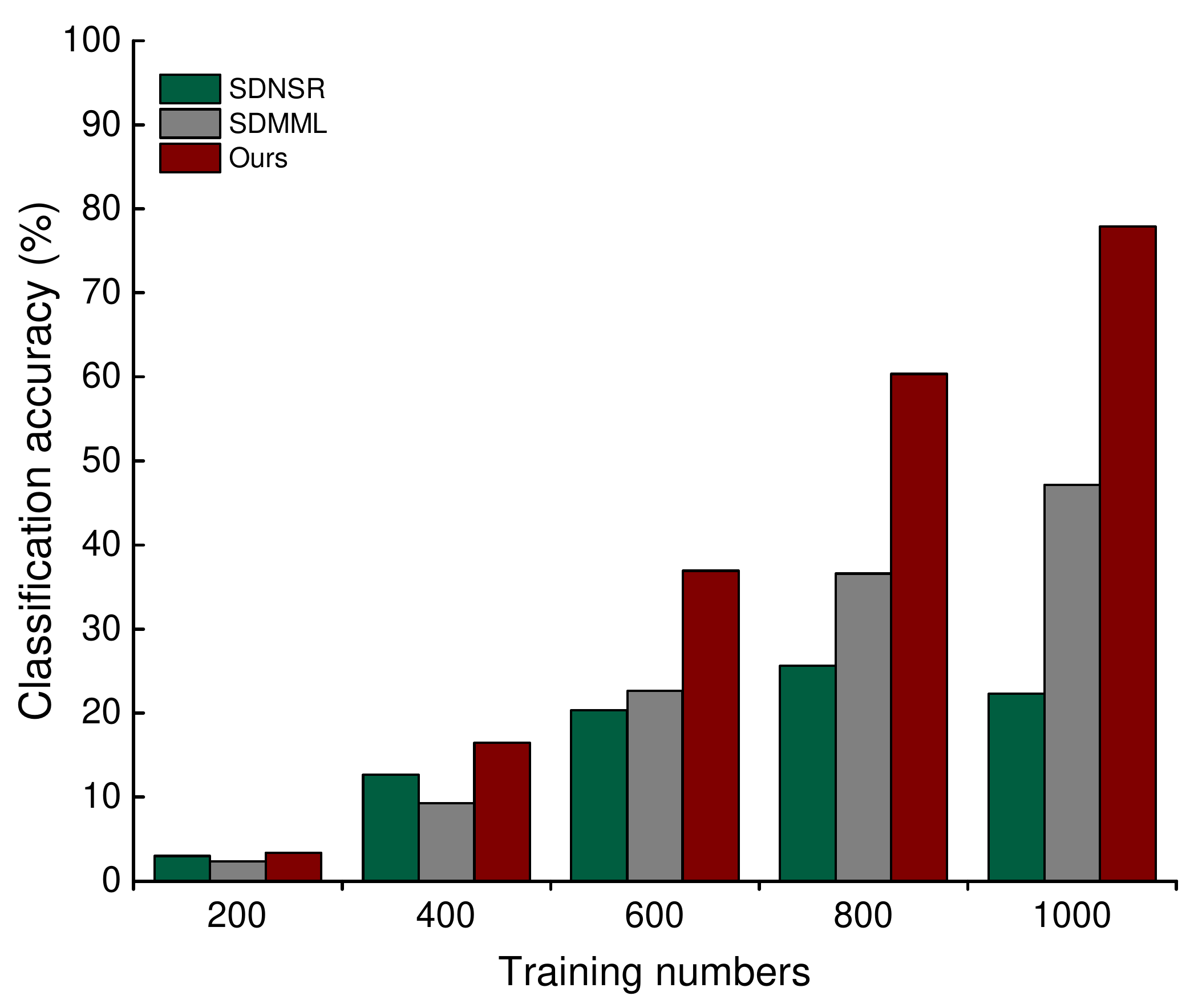}}
	\subfloat[]{%
		\includegraphics[width=1.7in, height=1.7in]{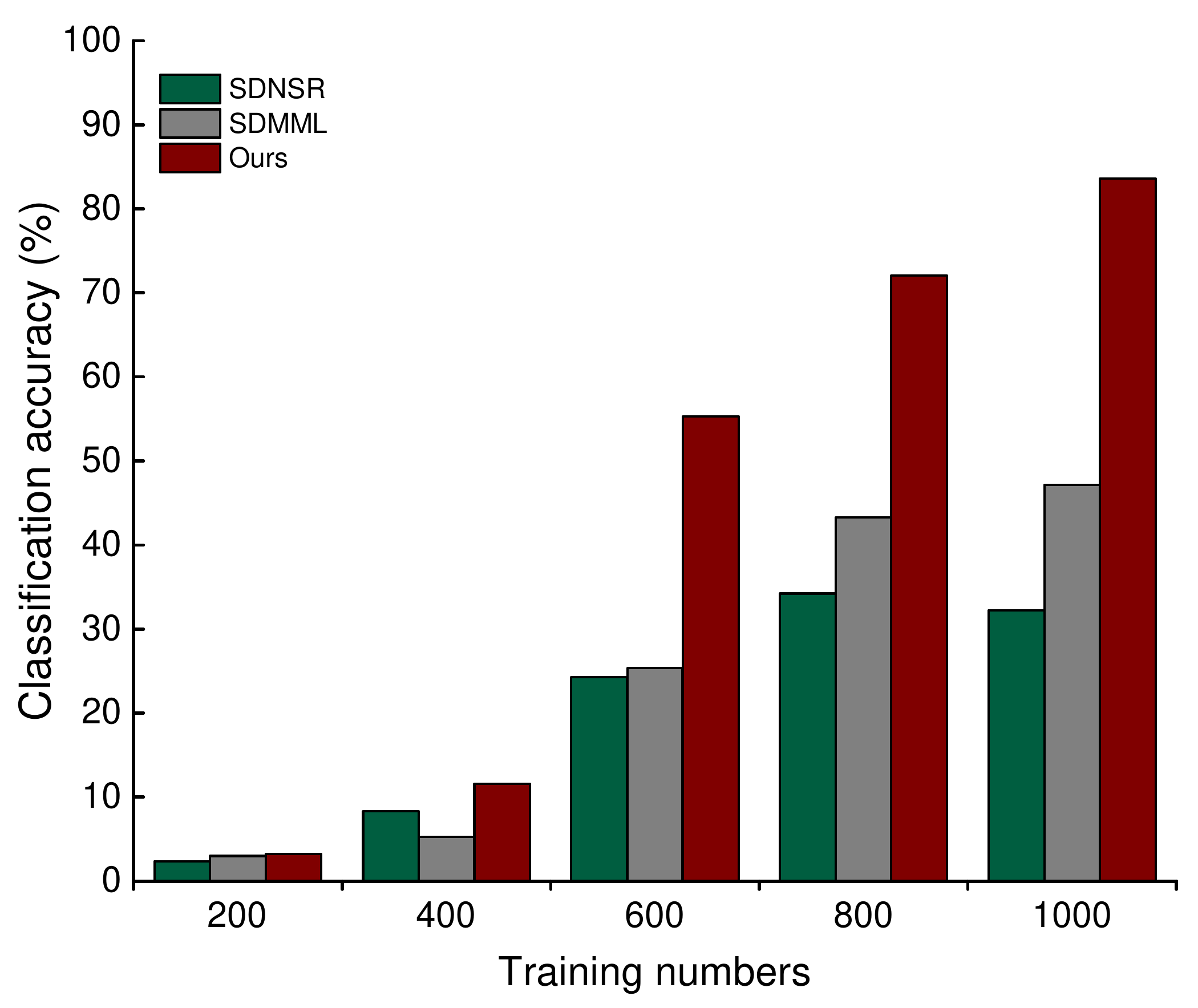}}
	\subfloat[]{%
		\includegraphics[width=1.7in, height=1.7in]{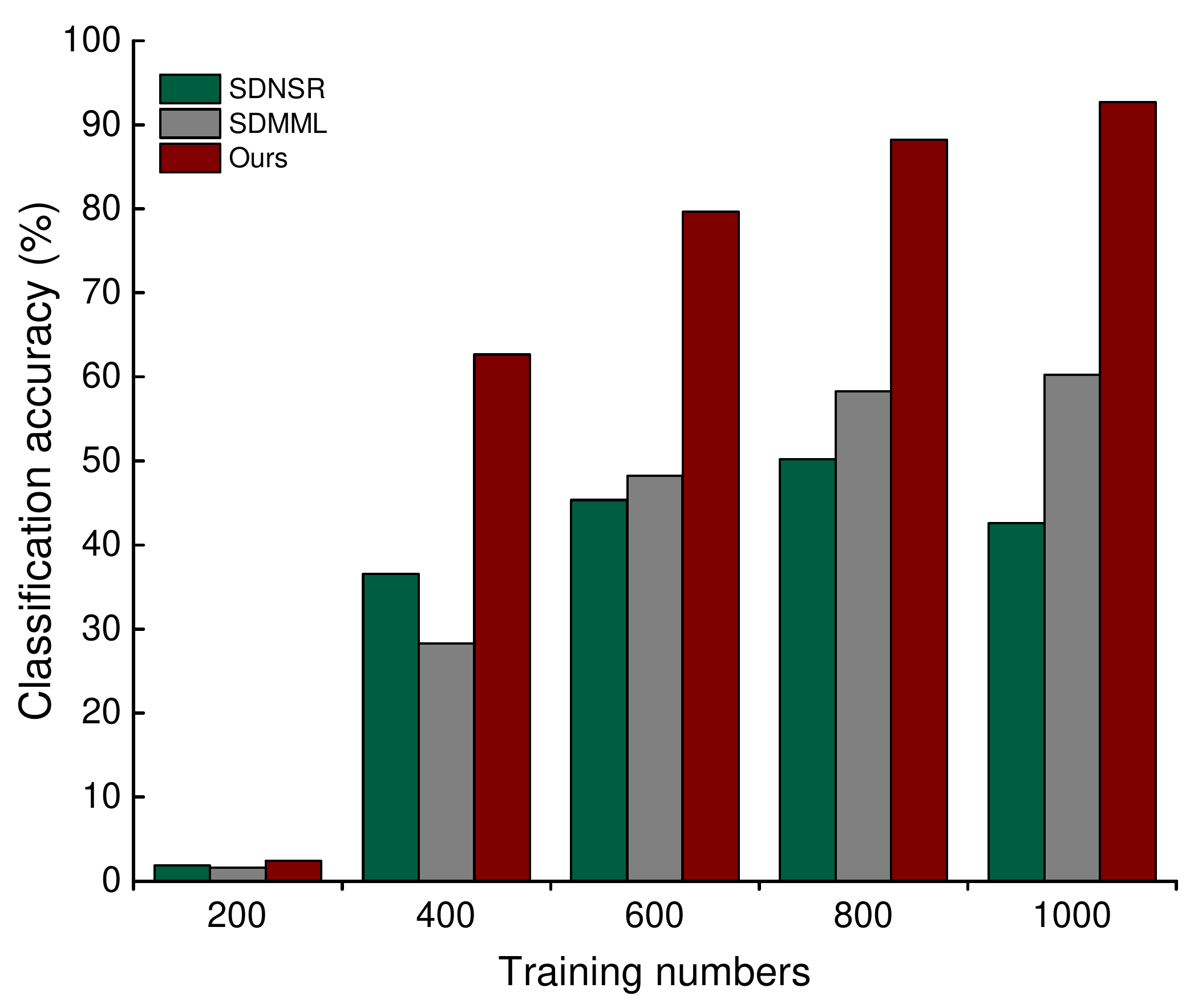}}
	\caption{Image classification ability of SDNSR, SDMML and ours on PASCAL VOC 2007 dataset: (a) ``airplane", (b) ``bird", (c) ``dog", (d) ``horse".}
	\label{new_fig1}
\end{figure*} 

\begin{table}[htb]
	\centering
	\renewcommand{\arraystretch}{1.2}
	\caption{The detailed number of subcategories and images used for image sub-categorization in ImageNet.}
	\begin{tabular}{|p{3.05cm}<{\centering}|p{0.8cm}<{\centering}|p{0.60cm}<{\centering}|p{0.6cm}<{\centering}|p{0.6cm}<{\centering}|p{0.6cm}<{\centering}|}
		\hline		
		\textbf{Category}    & airplane & horse & bird & cat & dog  \\
		\hline	
		\textbf{Number of subcategories}      & 15       & 29    & 26   & 9   & 18   \\		
		\hline
		\textbf{Number of images}             & 1434     & 1402  & 2126 & 1404& 1603   \\		
		\hline			
	\end{tabular}
	\label{tab_new2}
\end{table} 

\subsubsection{Baselines} We compare the image sub-categorization ability of our method with four baseline methods:

$\bullet$ multi-SVM \cite{weston1998multi}. For multi-SVM method, the class number is 97. We directly use the retrieved images from the image search engine as the positive samples to learn classifiers.

$\bullet$ Sub-Cate \cite{hoai2013discriminative}. Method Sub-Cate takes joint clustering and classification for subcategories discovering. For this experiment, the latent cluster number for each parent category is known and equal to the number of given subcategories. 

$\bullet$ RNCMF \cite{ristin2015categories}. For RNCMF method, the labeled training data is unavailable for both ``coarse" (parent) categories and ``fine" (sub) categories. The training images are retrieved from the image search engine which may include noise due to the error index of image search engine. We assume there are five trees corresponding to our five parent categories and start the recursively learning. The depth of the tree for this experiment is all limited to two levels. 

$\bullet$ MMDL \cite{wang2013max}. MMDL formulate image selection as a multi-instance learning problem. For this experiment, the subcategories are assumed as ``bags" and the retrieved images therein as instances. We take the proposed multi-instance learning function to select images from the retrieved images and learn the image classifiers.

\subsubsection{Experimental results}

Fig. \ref{fig4} and Fig. \ref{fig5} present the image sub-categorization results achieved by different methods when using a varying number of training images and testing images. The accuracy is measured by the average classification rate per subcategory. 

By observing Fig. \ref{fig4}, the best performance is achieved by our method, which produces significant improvements over method Sub-Cate and multi-SVM, particularly the number of training images over 20 for each subcategory. The reason is our method considers the noise during the process of classifier learning. Due to the error index of the image search engine, some noise may be included. We need to select useful images from the retrieved candidate images to learn robust classifiers for each subcategory.

From Fig. \ref{fig4}, we notice that the performance of the multi-SVM and Sub-Cate peaks at the value of training numbers 20 or 30 and decreases monotonically after this peaks. One possible explanation is that the image search engine provides images based on the estimated relevancy with respect to the query. Images far down in the ranking list are more likely to be noise, which may result in degrading of the sub-categorization accuracy especially for non-robust methods multi-SVM and Sub-Cate.

It is interesting to note in Fig. \ref{fig4}, while method RNCMF implements a form of noisy images removing, the classification accuracy did not improve with the number of training images increase. One possible explanation is that the noise in the training data is not the only factor that affects the classification accuracy. The visual distribution of the selected images is another important factor. Furthermore, the poor accuracy of Sub-Cate suggests that naively adding the number of training images without considering the visual distributions not only does not help but actually worsens the classification accuracy.

By observing Fig. \ref{fig4} and Fig. \ref{fig5}, our approach compares very favorably with competing algorithms, in terms of different numbers of training and testing images. Compared to method multi-SVM, Sub-Cate and RNCMF, our approach achieves significant improvements in the sub-categorization accuracy. The reason is our approach not only considers the possible presence of noise in the web training data, but also tries to ensure the diversity of the selected images for classifier learning. Besides, our approach performs better than method MMDL. The reason is we formulate image selection and classifier learning as a multi-class MIL problem. Compared to method MMDL which uses the relationship between different subcategories for classifier learning, our method not only uses the relationships between various subcategories, but also leverage the relationships between various parent categories. Therefore, our method achieves a much better result than MMDL.

\subsection{Quantitative Analysis of Different Steps}

Our proposed framework involves three major steps: subcategories discovering, noisy subcategories removing and multi-class multi-instance learning. In order to quantify the role of different steps contributing to the final classifiers, we construct two new frameworks. 
\begin{figure} [t]
	\centering
	\subfloat[]{%
		\includegraphics[width=1.7in, height=1.6in]{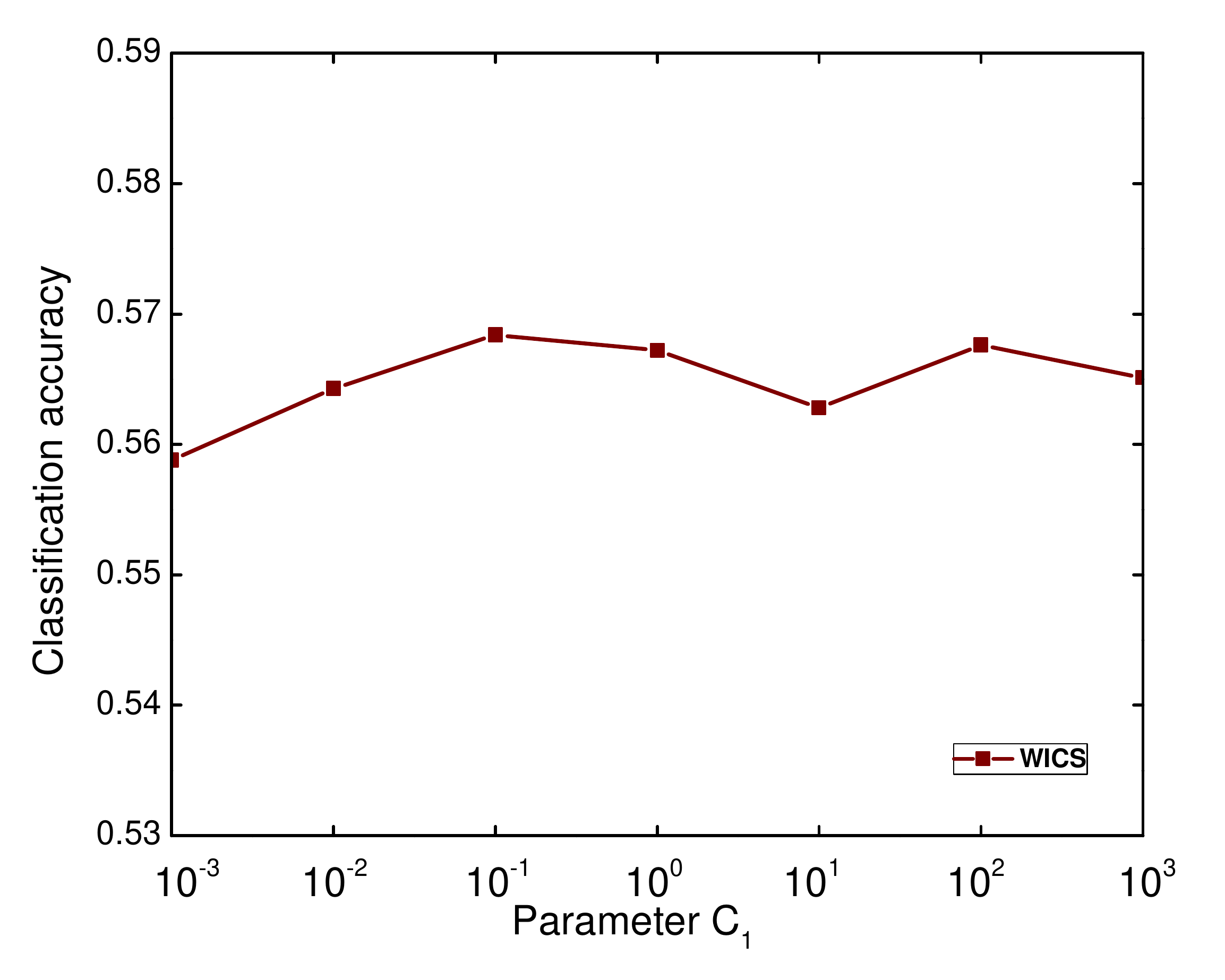}}
	\subfloat[]{%
		\includegraphics[width=1.7in, height=1.6in]{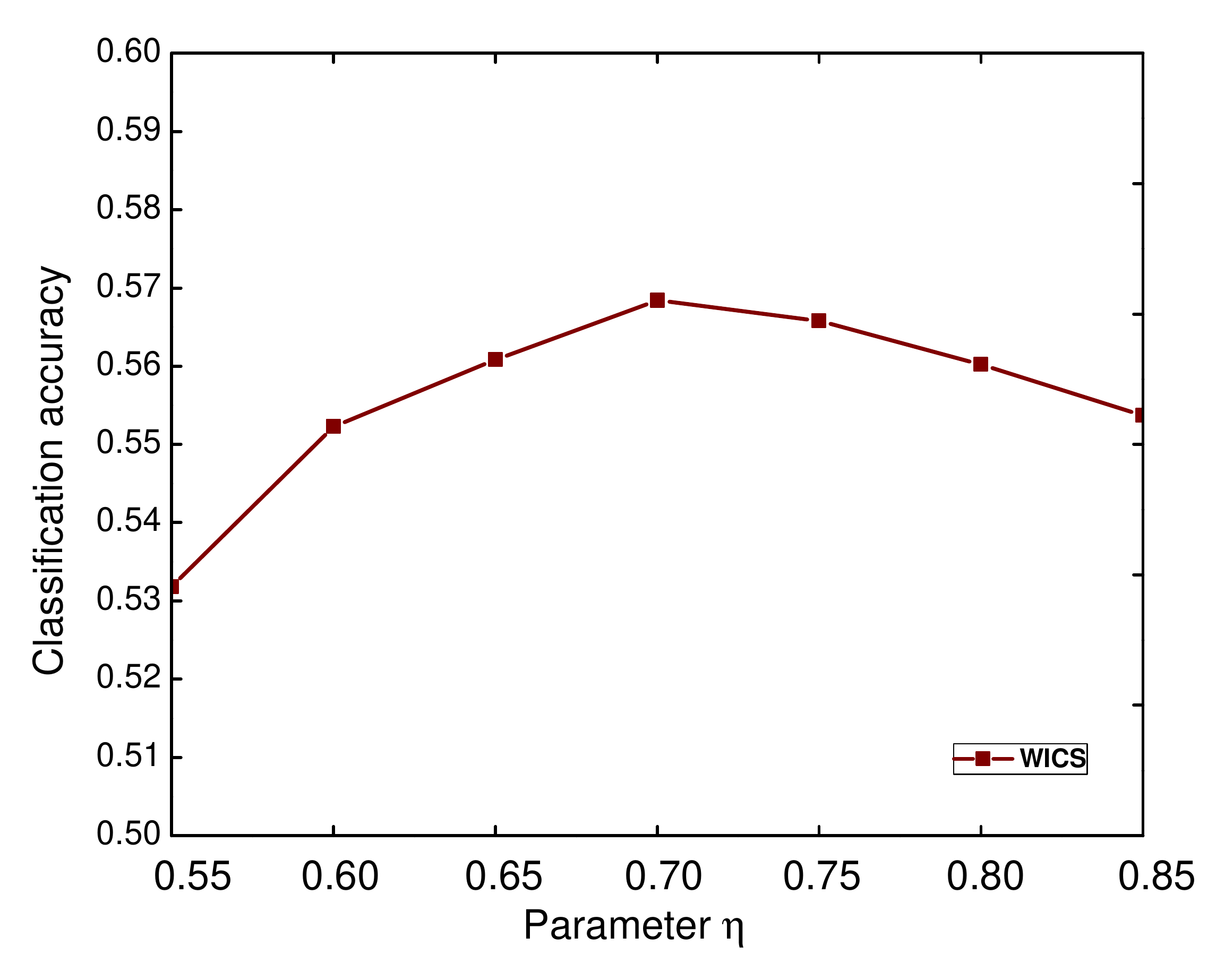}}
	\caption{The parameter sensitiveness of $ C_{1} $ and $\eta$ in terms of image classification accuracy.}
	\label{fig6}
\end{figure}
One is based on \textit{subcategories discovering and noisy subcategories removing} (which we refer to SDNSR).  Another one is based on \textit{subcategories discovering and multi-class multi-instance learning} (which we refer to SDMML). For framework SDNSR, we first obtain the candidate subcategories through searching in the Google Books Ngram Corpus. Then we apply the noisy subcategories removing procedure to get the selected subcategories. We directly retrieve the top images from the image search engine for selected subcategories to train image classifiers (without noisy images removing). For framework SDMML, we also obtain the candidate subcategories by searching in the Google Books Ngram Corpus. Then we retrieve the top images from the image search engine for all the candidate subcategories (without noisy subcategories removing). We apply the multi-class MIL model to select useful images and train image classifiers. 

We compare the image classification ability of these two new frameworks with our proposed framework. Specifically, ``airplane", ``dog", ``horse" and ``bird" are selected as four target categories to compare the image categorization ability. We sequentially collect [200,400,600,800,1000] images for each category as the positive training samples and use 1000 fixed irrelevant negative samples to learn image classifiers. We test the image classification ability of these three frameworks on the PASCAL VOC 2007 dataset. The results are shown in Fig. \ref{new_fig1}. By observing Fig. \ref{new_fig1}, we have the following observations: 

Framework SDNSR usually performs better than SDMML when the number of training images for each category is below 600. One possible explanation is that the first few images retrieved from the image search engine tend to have a relatively high accuracy. When the number of training images is below 600, the noisy images induced by noisy subcategories are more serious than those caused by the image search engine. With the increase of image numbers for each category, the images retrieved from the image search engine contain more and more noise. In this condition, the noisy images caused by the image search engine have a worse effect than those induced by noisy subcategories. 

Our proposed framework outperforms both SDNSR and SDMML. This is because our approach, which takes a combination of noisy subcategories removing and noisy images filtering, can effectively remove the noisy images induced by both noisy subcategories and the error index of image search engine. Our framework can maximize the filtering of noisy 
\begin{figure} [t]
	\centering
	\subfloat[]{%
		\includegraphics[width=1.7in, height=1.6in]{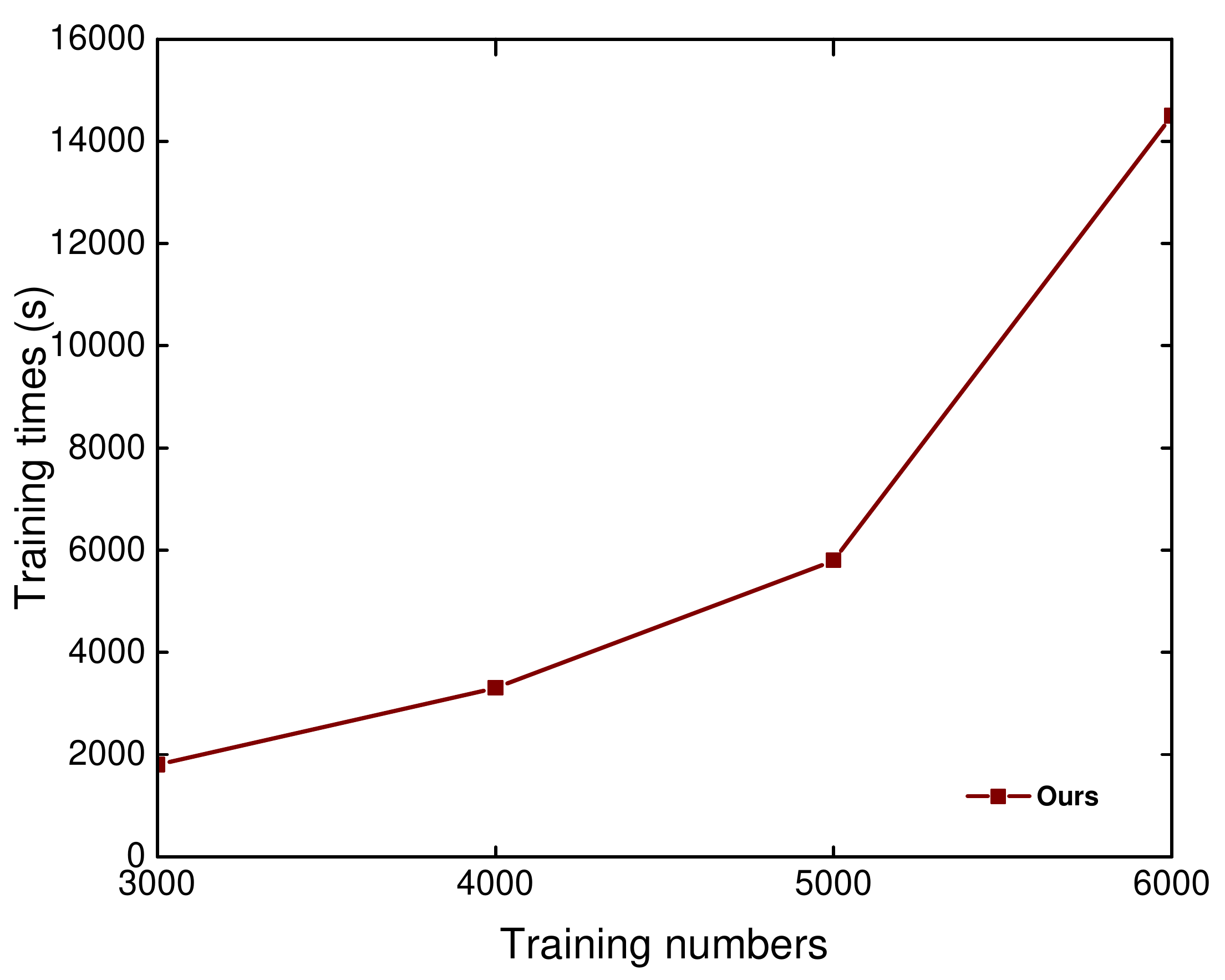}}
	\subfloat[]{%
		\includegraphics[width=1.7in, height=1.6in]{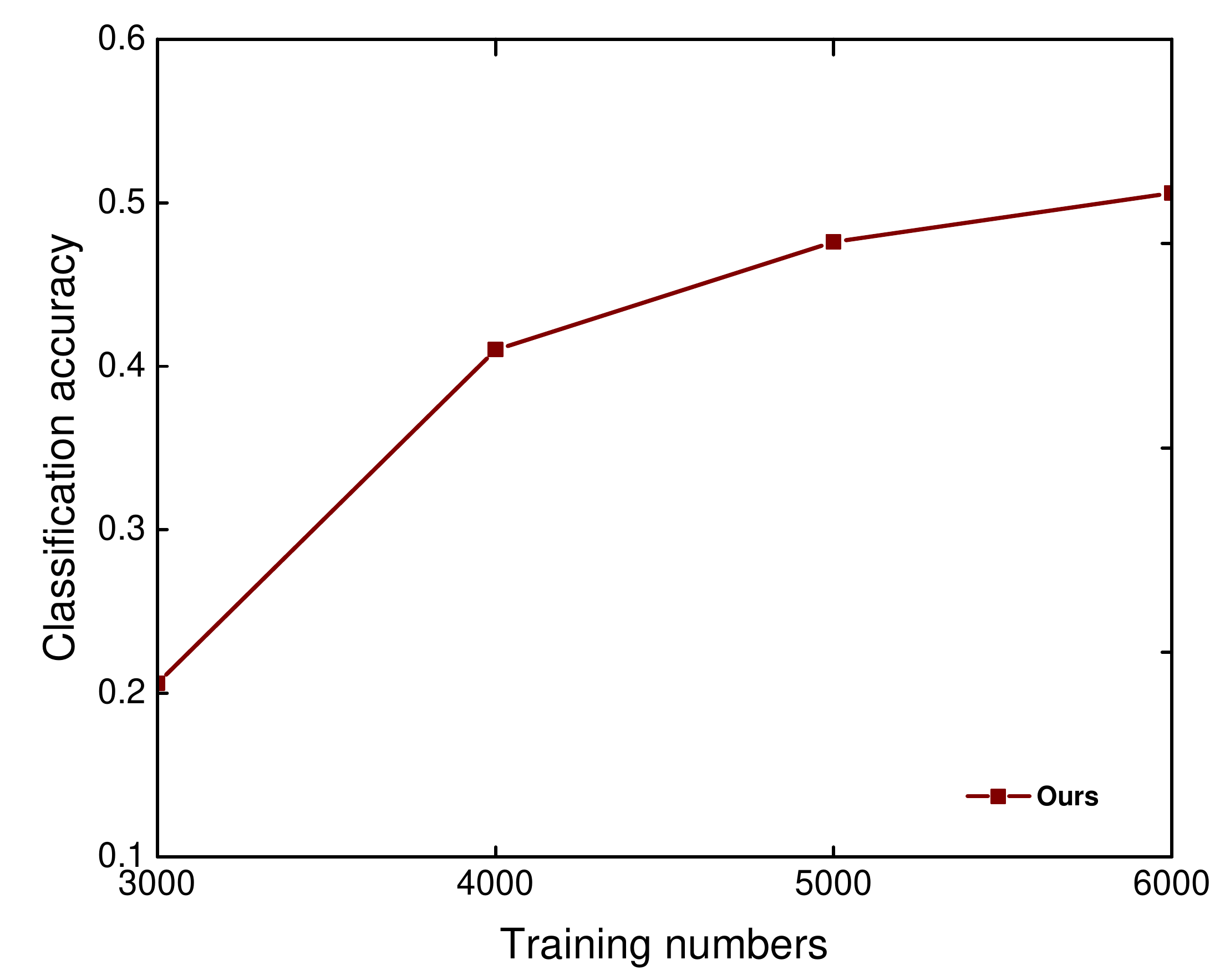}}
	\caption{The training time and image classification accuracies with respect to various numbers of training images.}
	\label{fig7}
\end{figure}
images while maintaining the diversity of the selected images, thereby reducing the negative impact of noisy images on the classifier.  

\subsection{Parameter Sensitivity Analysis} 

Our proposed multi-class multi-instance learning formulation contains two parameters $ C_{1} $ and $ \eta $. PASCAL VOC 2007 was selected as the benchmark testing dataset to evaluate the performance variation of our proposed approach. In particular, we vary one parameter by fixing another parameter as the default value. Fig. \ref{fig6} presents the parameter sensitiveness of $ C_{1} $ and $\eta$ in terms of image classification accuracy. 

By observing Fig. \ref{fig6}, we found our method is robust to the parameter $ C_{1} $ when it is varied in a certain range. Besides, the performance of our 
method is growing when $ \eta $ increase but less than 0.7. The reason is perhaps that our training data derived from image search engine. Due to the error index of image search engine, there may be too much noise in each bag which will result in decreasing the classification accuracy when $ \eta \leqslant 0.7 $. When $ \eta $ increases over 0.7, the performance of our method decreases. One possible explanation is that the training set is less diverse. With the increasing of $ \eta $, the number of subcategories is decreasing, which may lead to the degradation of domain robustness of the classifier.  

\subsection{Time Complexity Analysis}

During the process of multi-class multi-instance learning, we solve the convex problem in \eqref{eq12} by the cutting-plane algorithm. Through finding the most violating candidate $ \textbf{h}_{t} $ and solve the MKL subproblem at each iteration, the time complexity of \eqref{eq12} can be approximately computed as $ T\cdot O $(MKL), where the \textit{T} is the number of iterations and the $ O $(MKL) is the time complexity of the MKL sub-problem. According to \cite{platt199912,niu2016visual}, the time complexity of MKL is between $ t \cdot O(LCM) $ and $ t \cdot O((LCM)^{2.3}) $, where $ M, L, C  $ are the numbers of latent domains, bags and categories respectively. $ t $ is the number of iterations in MKL. 

We take STL-10 as the testing set to evaluate our method. Particularly, we use various numbers of training images for each category to learn the classifiers. STL-10 has 10 categories and we use $ n $ training images for each category, so we have a total of 10 $ n $ training images. Fig. \ref{fig7} shows the training time and image classification accuracies with respect to various numbers of training images. From Fig. \ref{fig7}, we can observe that both of training time and image classification accuracy increase with the number of training images grows. We report the configuration of our experiment. Two HP PCs (3.2GHz CPU with 8 GByte RAM) were used for the web images collection. All the data processing and experiments were performed on an Acer PC (3.5GHz CPU, 16 GByte RAM and 4 GByte VRAM) with LIBSVM \cite{chang2011libsvm}.

\section{Conclusion} 

In this paper, we presented a new framework for classifying images into categories and subcategories. Three successive modules were employed in the framework including subcategories discovering, noisy subcategories removing and multi-class multi-instance learning. Compared to existing methods, our proposed approach can not only classify images into NOUN  subcategories, but also into VERB, ADJECTIVE and ADVERB subcategories. Our approach has a better semantic refinement descriptions for the categories. To verify the effectiveness of our proposed approach, we conducted experiments on both image categorization and sub-categorization tasks. The experimental results demonstrated the superiority of our proposed approach over existing weakly supervised and web-supervised approaches.

\appendices

\section{THE DERIVATIONS OF \eqref{eq10} is the dual form of \eqref{eq3}}

\emph{Proof:} In order to deduce the dual form of \eqref{eq3}, we introduce an variable $\pi$ which is defined as:
$$
\pi _{i,c,s,\hat{s}}= \left\{\begin{matrix}
\hat{P_{i,s}} & c = y_i\\ 
\delta (s=\hat{s}) & c \neq y_i
\end{matrix}\right.
$$
where $y_i = Y_m, \forall i \in I_m$. Then, we can get:
$$
\sum_{s=1}^{S}\hat{P_{i,s}}{\mathbf{w}_{y_i,s}^\top}\phi (\textbf{x}_i) = \sum_{s=1}^{S}\pi _{i,y_i,s,\hat{s}}{\mathbf{w}_{y_i,s}^\top}\phi (\textbf{x}_i) 
$$
$$
{\mathbf{w}_{c,\hat{s}}^\top}\phi (\textbf{x}_i) = \sum_{s=1}^{S}\pi _{i,c,s,\hat{s}}{\mathbf{w}_{c,s}^\top}\phi (\textbf{x}_i).
$$
We can further rewrite the constraints in \eqref{eq4} as the following forms: 
$$
\frac{1}{\left |G_m  \right |}\sum _{i\in I_m} h_i(\sum_{s=1}^{S}\pi_{i,Y_m,s,\hat{s}}{\left (\mathbf{w}_{Y_m,s}  \right )}^\top\phi (x_i)-\sum_{s=1}^{S}\pi_{i,c,s,\hat{s}}$$
$$ {\left (\mathbf{w}_{c,s}  \right )}^\top\phi (x_i)) \geq \zeta _{m,c,\hat{s}}-\xi _m,\; \forall m,\hat{s},c
$$ 
where $\zeta _{m,c,s}=0$ if $c=Y_m$ and $\zeta _{m,c,s}=\eta $ otherwise. We define:
$$\mathbf{w}={\left [ {\mathbf{w}}^\top_{1,1},...,{\mathbf{w}}^\top_{1,S},{\mathbf{w}}^\top_{2,1},...,{\mathbf{w}}^\top_{C,S} \right ]}^\top$$
and a new mapping function for $G_m$ as:
$$
\psi (\mathbf{h},G_m,c,\hat{s})=\left(\frac{1}{\left |G_m  \right |}\sum _{i\in I_m} h_i\pi_{i,1,1,\hat{s}}\delta (c=1){\phi (x_i)}^\top,..., \right.  \\ $$
$$
\left. \frac{1}{\left |G_m  \right |}\sum _{i\in I_m} h_i\pi_{i,C,S,\hat{s}}\delta (c=C){\phi (x_i)}^\top  \right)^\top.
$$
By further denoting:
$$
\varphi(\mathbf{h},G_m,c,\hat{s})=\psi (\mathbf{h},G_m,Y_m,\hat{s})-\psi (\mathbf{h},G_m,c,\hat{s}),
$$
the problem of \eqref{eq3} can be written as: 
\begin{equation}\label{new18}
\begin{aligned}
\min_{\textbf{h},\mathbf{w},\xi _m}\frac{1}{2}\left \| \mathbf{w} \right \|^{2}+C_{1}\sum_{m=1}^{M}\xi _m \\
\end{aligned}
\end{equation}
\begin{equation}\label{new19}
\begin{aligned}
\mathrm{s.t.}\; {\mathbf{w}}^\top\varphi (\mathbf{h},G_m,c,s)\geq \zeta _{m,c,s}-\xi _m,\; \; \forall m,c,s.  
\end{aligned}
\end{equation}

By introducing a dual variable $\alpha _{m,c,s}$ for each constraint in \eqref{new19}, we can get the Lagrangian as:
\begin{equation}\label{new20}
\begin{aligned}
\pounds_{\mathbf{w},\xi _m,\alpha _{m,c,s}}=\frac{1}{2}\left \| \mathbf{w} \right \|^{2}+C_{1}\sum_{i=1}^{N}\xi _{i}- \\
\sum _{m,c,s} \alpha _{m,c,s}({\mathbf{w}}^\top\varphi (\mathbf{h},G_m,c,s) - \zeta _{m,c,s}+\xi _m).
\end{aligned}
\end{equation}
Through set the derivatives of $\pounds$ with respect to $\mathbf{w}$ and $\xi _m$ as zeros, we can get:
\begin{equation}\label{new21}
\begin{aligned}
\sum_{c,s} \alpha_{m,c,s} = C_{1}
\end{aligned}
\end{equation}
\begin{equation}\label{new22}
\begin{aligned}
\mathbf{w} = \sum_{m,c,s} \alpha_{m,c,s} \varphi (\mathbf{h},G_m,c,s)
\end{aligned}
\end{equation}

Through submit the obtained equalities \eqref{new21} and \eqref{new22} back into \eqref{new20}, we can get the dual form of \eqref{eq3} as \eqref{eq10}, which completes the proof.

\section{THE DERIVATIONS OF \eqref{eq11} is the dual form of \eqref{eq12}}

\emph{Proof:} We firstly introduce a dual variable $\alpha_{m,c,s}$ for constraint in \eqref{eq13}, then we can rewrite the Lagrangian form of \eqref{eq12} as:
\begin{equation}\label{eq14}
\begin{aligned}
\pounds  =\frac{1}{2}\sum_{t=1}^{T}\frac{\left \| \mathbf{\mathrm{w}}_{t} \right \|^2}{d_t}+C_1\sum_{m=1}^{M}\xi_m -\sum _{m,c,s}\alpha_{m,c,s} \\
(\sum_{t=1}^{T}\mathrm{\mathbf{w}}_t^{\top}\varphi (\mathrm{\mathbf{h}}_t,G_m,c,s)- \zeta _{m,c,s}+\xi _m) \\
\end{aligned}
\end{equation}
Through set the derivatives of $\pounds$ w.r.t. $\mathbf{w}_t$ and $\xi _m$ as zeros respectively, we can get:
\begin{equation}\label{eq15}
\begin{aligned}
\mathbf{w}_t=d_t\sum _{m,c,s}\alpha _{m,c,s}\varphi (\mathbf{h}_t,G_m,c,s),\: \forall t, \\
\sum _{c,s}\alpha _{m,c,s}=C_1,\: \forall m.
\end{aligned}
\end{equation}
Through submit \eqref{eq15} back into \eqref{eq12}, we can arrive at the object function in \eqref{eq11}, which completes the proof.


\ifCLASSOPTIONcaptionsoff
\newpage
\fi

\end{document}